%% file: panscales-soft-spin.tex
\documentclass[a4paper,11pt]{article}
\pdfoutput=1 
\RequirePackage{snapshot}  

\usepackage{jheppub} 

\usepackage{bm,amsmath,amssymb,slashed,graphicx,%
            enumerate,alltt,xspace,multirow,xcolor,mathrsfs}
\usepackage{fancyvrb}
\usepackage{booktabs,tabularx}
\usepackage{graphicx}
\usepackage{subcaption}
\usepackage{xspace}
\usepackage[utf8]{inputenc} 
\usepackage[compat=1.0.0]{tikz-feynman}
\usepackage{scalerel}
\usepackage{layouts}

\usepackage{printlen}
\usepackage{wasysym}

\usepackage[capitalise]{cleveref}
\makeatletter
\g@addto@macro\bfseries{\boldmath}
\makeatother

\definecolor{labelkey}{rgb}{0,0.5,0.0}
\definecolor{royalpurple}{rgb}{0.47, 0.32, 0.66}

\usepackage{listings}
\definecolor{darkgreen}{rgb}{0,0.4,0}
\definecolor{grey}{rgb}{0.5,0.5,0.5}

\allowdisplaybreaks 

\definecolor{semiblue}{rgb}{0.3,0.3,0.8}
\newcommand{\logbook}[2]{}

\newcommand{\order}[1]{\mathcal{O}\left(#1\right)}
\newcommand{\as}{\alpha_s}

\newcommand{\ptilde}{{\widetilde p}}

\newcommand{\nc}{N_\text{\textsc{c}}}

\newcommand{\itilde}{{\tilde \imath}}
\newcommand{\jtilde}{{\tilde \jmath}}

\newcommand{\zcut}{z_{\text{cut}}}

\newcommand{\dpsisl}{\Delta\psi_{12}^{\mathrm{slice}}}

\tikzstyle{block} = [rectangle, minimum width=1.0cm, minimum height=0.75cm, thin, draw=black]
\tikzstyle{blob} = [circle, minimum width=0.5cm, thin, draw=black]
\tikzset{blackarrow/.style={-stealth, semithick, draw=black}}
\tikzset{connection/.style={inner sep=0,outer sep=0}}

\newcolumntype{C}{>{\centering\arraybackslash}X}

\title{Soft spin correlations in final-state parton showers}

\preprint{OUTP-21-25P}

\newcommand{\OXaff}{Rudolf Peierls Centre for Theoretical Physics, Clarendon Laboratory, Parks Road,
  University of Oxford, Oxford OX1 3PU, UK}
\newcommand{\ASCaff}{All Souls College, Oxford OX1 4AL, UK}

\author[a]{Keith~Hamilton,}%
\author[b]{Alexander Karlberg,}%
\author[b,c]{Gavin~P.~Salam,}%
\author[b]{Ludovic Scyboz,}%
\author[a]{Rob Verheyen}%

\affiliation[a]{Department of Physics and Astronomy, University College London, London, WC1E 6BT, UK}
\affiliation[b]{\OXaff}
\affiliation[c]{\ASCaff}

\date{Received: date / Accepted: \today}

\abstract{
  We introduce a simple procedure that resolves the long-standing
  question of how to account for single-logarithmic spin-correlation effects in parton showers
  not just in the collinear limit, but also in the soft wide-angle
  limit, at leading colour.
  We discuss its implementation in the context of the PanScales family
  of parton showers, where it complements our earlier treatment of 
  the purely collinear spin correlations.
  Comparisons to fixed-order matrix elements help validate our
  approach up to third order in the strong coupling, and an appendix
  demonstrates the small size of residual
  subleading-colour effects.
  To help probe wide-angle soft spin correlation effects, we introduce
  a new declustering-based non-global spin-sensitive observable, the
  first of its kind.
  Our showers provide a reference for its single-logarithmic
  resummation.
  The work in this paper represents the last step required for
  final-state massless showers to satisfy the broad PanScales next-to-leading logarithmic
  accuracy goals.
} 

\keywords{QCD, Parton Shower, Resummation, LHC, LEP}

\begin{document}

\maketitle

\section{Introduction}
\label{sec:intro}

Parton showers are some of the most extensively used tools in collider
physics.
For much of the past decades the main aim for parton showers, from the
point of view of resummation, was to achieve leading-logarithmic (LL)
accuracy, i.e.\ control of terms $\as^n L^{2n}$ or in some cases
$\as^n L^{n+1}$, where $\as$ is the strong coupling and $L$ is the
logarithm of a ratio of some pair of disparate momentum scales.
In recent years there have been a number of advances in formulating
showers with well-defined resummation
accuracy~\cite{Dasgupta:2018nvj,Dasgupta:2020fwr,Hamilton:2020rcu,Karlberg:2021kwr,Forshaw:2020wrq,Holguin:2020joq,Nagy:2020rmk,Nagy:2020dvz},
making the prospect of an NLL-accurate shower a concrete possibility.
The identification of the logarithmic accuracy of tools such as a
parton shower, that can be used to predict arbitrarily complex
observables, is not without subtleties.
One comprehensive proposal for a definition of the logarithmic
accuracy was given in Refs.~\cite{Dasgupta:2018nvj,Dasgupta:2020fwr}.
Within that proposal, to claim NLL accuracy, a shower should correctly
reproduce all sources of potential $\as^n L^{n}$ contributions, at
least at leading colour.\footnote{While full colour is highly
  desirable at LL accuracy, one may argue that it is legitimate to
  leave out subleading-colour effects in $\as^n L^{n}$ terms, because
  their suppression by powers of $1/\nc^2$, where $\nc =3$ is the
  number of colours, renders them quantitatively similar to
  $\as^n L^{n-1}$ (NNLL) contributions.}
It should also reproduce tree-level matrix elements in the limits
where all branchings are well separated in the Lund
diagram~\cite{Andersson:1988gp} (these are the most common
configurations and so, arguably, the configurations most likely to
be targeted by machine-learning based jet-tagging
methods~\cite{Larkoski:2017jix}).

One fundamental set of quantum mechanical effects that has long been
known to play a role in showering is that of spin
correlations~\cite{Webber:1986mc,Collins:1987cp,Knowles:1987cu}.
Insofar as parton showers aimed only for LL accuracy, the inclusion of
spin correlations could be considered optional and of the major parton
shower codes, only the Herwig\footnote{There has also been work on
  collinear spin correlations in other shower frameworks, see
  Refs.~\cite{Nagy:2007ty,Nagy:2008eq,Forshaw:2019ver,Forshaw:2020wrq}.}
showers include them fully to all orders in the collinear
limit~\cite{Corcella:2000bw,Bahr:2008pv,Bellm:2015jjp,Richardson:2018pvo,Webster:2019cwq}.
However, starting from NLL accuracy, the inclusion of spin
correlations is no longer optional.

\begin{figure}
\centering
  \begin{subfigure}{0.49\textwidth}
  \includegraphics{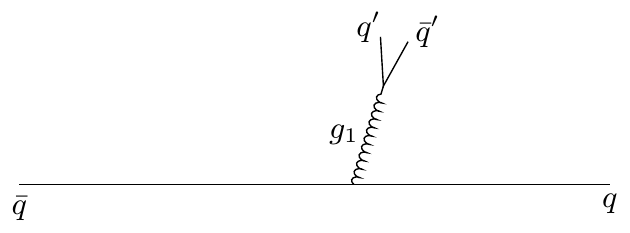}
  \includegraphics{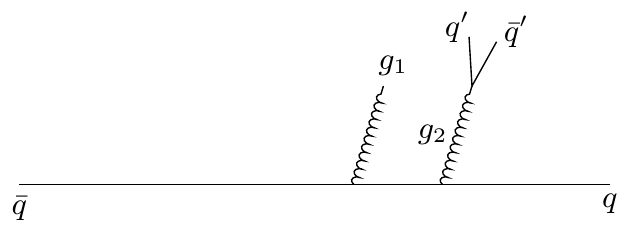}
  \end{subfigure}
  \begin{subfigure}{0.49\textwidth}
  \includegraphics{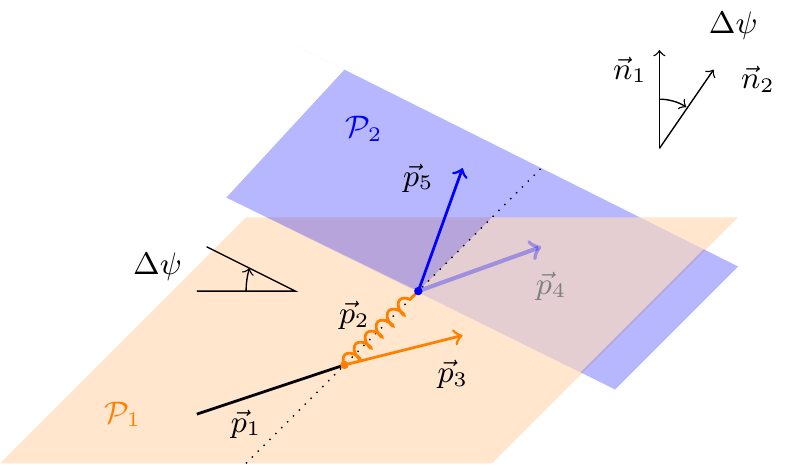}
  \end{subfigure}
  \caption{Left:~Configurations at $\mathcal{O}(\alpha_s^2)$ (top) and
    $\mathcal{O}(\alpha_s^3)$ (bottom), where a soft, wide-angle gluon
    ($E_{g_2} \ll E_{g_1} \ll E_{q,\bar q}$ and
    $\theta_{g_1q}\sim \theta_{g_2q} \sim \order{1}$) splits
    collinearly ($\theta_{q'\bar q'} \ll 1$).
    This generates spin correlations that are sensitive to soft
    corrections of the type discussed in
    \cref{sec:technical}.
    Right:~Illustration of the definition of the spin-sensitive azimuthal
    difference $\Delta \psi$ between subsequent branching planes.
  }
\label{fig:results-alphas-conf}
\end{figure}

Many aspects of spin correlations can be accounted for with the
Collins algorithm~\cite{Collins:1987cp}, which provides an efficient
and straightforward approach for the spin correlations in collinear
splittings.
However, parton showers can involve interleaved sequences of soft,
large-angle splittings and collinear splittings, and these bring in
spin correlations that are beyond the scope of the Collins algorithm.
As an example, \cref{fig:results-alphas-conf} (left) shows configurations
that contribute to single-logarithmic terms, at second order (top)
and third order (bottom). In the former case, we consider the emission
of a soft gluon (relative to the Born quark-pair),
$ E_{g_1} \ll E_{q},E_{\bar q} $, at large angle
$\theta_{g_1} \sim 1$, where that gluon splits collinearly to either
$g\to gg$ or $g\to q' \bar q'$ (in the figure,
$\theta_{q' \bar q'} \ll 1$).
At one order higher, we consider two
gluons that are ordered in energy,
$ E_{g_2} \ll E_{g_1} \ll E_{q},E_{\bar q}$, but at commensurate, large
angles, $\theta_{g_1} \approx \theta_{g_2} \sim 1$ with respect to the
Born quark-pair, followed by the collinear splitting $g_2 \to gg$, or
$g_2 \to q' \bar q'$.
As in the case just of collinear splittings, the
spin carried by the intermediate gluon will induce a correlation of
the azimuthal angles between the successive splitting
planes.
This is illustrated in \cref{fig:results-alphas-conf} (right).

In this article, we introduce a simple extension of the Collins
algorithm, relevant for dipole showers, that enables it to
simultaneously address the collinear and the soft large-angle regimes, providing an
efficient solution to the last remaining barrier to obtaining an
NLL-accurate massless final-state shower at leading-$\nc$
(\cref{sec:technical}).
\cref{sec:fo-validation} presents our tests of the algorithm against
fixed-order matrix elements.
In \cref{sec:new-observable} we introduce a new observable that is
sensitive to the pattern of large-angle soft emissions as well as the
azimuthal structure of the subsequent collinear splittings of those
large-angle soft emissions.  We conclude in \cref{sec:concl}. A number
of appendices are included which provide more details on certain
aspects of our work. In \cref{app:analytic-mes} we give explicit
expressions for the 4- and 5-parton soft-limit matrix elements used in
our validation. \cref{sec:app:FC} discusses subleading-$\nc$ effects
and \cref{sec:app:reference-direction} investigates the systematic
effects associated with the choice of reference vector in the evaluation of
spinor products.

\section{Extending the Collins algorithm to the wide-angle soft limit}
\label{sec:technical}
\begin{table}
  \centering
  \begin{tabular}{  c | c | c | c | c | c  }
  \toprule
  $\lambda_\itilde$  & $\lambda_i$  & $\lambda_k$  & $q \to qg$ & $g \to q\bar{q}$ & $g \to gg$    \\
  \midrule
  $\lambda$ &      $\lambda$        &      $\lambda$       & $\frac{1}{\sqrt{1-z}}$ & $0$ & $\frac{1}{\sqrt{z(1-z)}}$ \\
  $\lambda$ &      $\lambda$        &      $-\lambda$      & $\frac{z}{\sqrt{1-z}}$ & $-z$ & $\frac{z^{3/2}}{\sqrt{1-z}}$ \\ 
  $\lambda$ &      $-\lambda$       &      $\lambda$       & $0$ & $1-z$ & $\frac{(1-z)^{3/2}}{\sqrt{z}}$ \\
  $\lambda$ &      $-\lambda$       &      $-\lambda$      &           $0$               &$0$                  &                 $0$    \\
  \end{tabular}
  \caption{The helicity-dependent Altarelli-Parisi splitting
    amplitudes $\mathcal{F}_{\itilde \to ik}^{\lambda_\itilde \lambda_i
      \lambda_k}(z)$, where $z = E_i/E_\itilde$.} 
\label{tab:hel-dep-fcns}
\end{table}

The standard procedure for including spin correlations in a parton
shower is the Collins--Knowles
algorithm~\cite{Collins:1987cp,Knowles:1987cu}.
It relies crucially on the construction of a binary tree of $1\to 2$
collinear splittings, with strongly ordered angles as one moves down
the tree, i.e.\ towards the shower's final set of particles.
The algorithm effectively maintains spin-density information across
all nodes of the tree, and for each new $1 \to 2$ splitting updates
that spin-density information at all of the ancestor nodes of the splitting node.
The difficulty that we face in including spin correlations for soft
large-angle emissions is that the corresponding splittings necessarily
involve a $2 \to 3$ structure.\footnote{This is in the large-$\nc$
  limit, while beyond that limit the structure becomes more
  complicated.}
As such they do not obviously mesh with the Collins--Knowles algorithm.
We address this problem within the adaptation of the Collins--Knowles
algorithm that we introduced in the PanScales framework in
Ref.~\cite{Karlberg:2021kwr}.\footnote{See also
  Refs.~\cite{Richardson:2018pvo,Webster:2019cwq} for an alternative
  dipole-shower approach.
  It differs sufficiently from the PanScales approach, that addressing
  soft emissions in there would require a dedicated study.}
It was intended for use in final-state dipole and antenna showers,
and we demonstrated that it correctly reproduces the spin-induced
azimuthal dependence of collinear radiation in single-logarithmic
terms, but did not explore the question of the soft arbitrary-angle
branching.\footnote{
  We use the terms ``wide'' and ``large'' angle
  interchangeably to denote angles of order $1$.
  The term ``arbitrary'' angles includes large angles, as well as
  angles that are commensurate with those of emissions that occurred
  earlier in the parton shower.
}

One of the key elements of the PanScales version of the
Collins--Knowles algorithm is that when an $n$-parton system
undergoes a collinear $\itilde \to ik$ splitting to give an
$(n+1)$-parton system, the colour-stripped amplitudes ($M$) for the
$(n+1)$- and $n$-parton systems can be related by collinear
factorisation,
\begin{equation}
  \label{eq:collinear-factorization}
  M^{\lambda_i\lambda_k}(\ldots,p_i,p_k,\ldots)
  = \mathcal{M}_{\itilde \to ik}^{\lambda_\itilde \lambda_i \lambda_k}
  \times
  M^{\lambda_\itilde}(\ldots,p_\itilde,\ldots)\,,
\end{equation}
where all helicity indices ($\lambda_1\ldots \lambda_n$) and momenta
($p_1 \ldots p_n$) that are not explicitly shown
remain the same on both sides.
The effective splitting amplitude $\mathcal{M}_{\itilde \to
  ik}^{\lambda_\itilde \lambda_i \lambda_k}$ is given by
\begin{equation}
\mathcal{M}_{\itilde \to ik}^{\lambda_\itilde \lambda_i \lambda_k} =
\frac{1}{\sqrt{2}} \frac{g_s}{p_i{\cdot}p_k} \mathcal{F}_{\itilde \to ik}^{\lambda_\itilde
\lambda_i \lambda_k}(z) S_{\tau}(p_i,p_k)\;.
\label{eq:branch-amp-spinor-prod}
\end{equation}
Here, $z$ is the momentum fraction of $i$ relative to $\itilde$, while 
$\lambda_a = \pm 1$ corresponds to the helicity of parton $a$. 
The spinor product $S_\tau(p_i,p_k)$ (which follows the conventions of
Ref.~\cite{Kleiss:1985yh}) involves a
complex phase that depends on the azimuth of the $ik$ angle and the
spin index $\tau=\pm1$, which is given by
\begin{equation}
\tau = \tilde{\lambda}_i + \tilde{\lambda}_k - \tilde{\lambda}_\itilde \mbox{ where } \tilde{\lambda} = 
\begin{cases}
  \lambda/2 \mbox{ for a quark},  \\ 
\lambda \mbox{ for a gluon}.
\end{cases}
\end{equation}
The interplay between the phases at different nodes of the
branching tree (summing over amplitude and complex-conjugate amplitude
spin indices at successive splittings) ultimately leads to azimuthal
correlations between splittings across the tree.
The functions
$\mathcal{F}_{\itilde \to ik}^{\lambda_\itilde \lambda_i
  \lambda_k}(z)$ are (real-valued) colour-stripped helicity-dependent
Altarelli-Parisi splitting amplitudes, which depend on the momentum
fraction $z$ carried by parton $i$, and are given in
\cref{tab:hel-dep-fcns}.  We refer the reader to Appendix A of
Ref.~\cite{Karlberg:2021kwr} for the derivation of
\cref{eq:branch-amp-spinor-prod} and further details on the
conventions used in the spinor products.

Next, let us look at the analogous amplitudes in the limit where we
allow the emitted gluon to be at arbitrary angles but require it to be
soft relative to its parents.
This involves a $2\to3$ structure, $\itilde \jtilde \to ijk$, and the
relation between the $n$- and $(n+1)$-parton colour-stripped amplitudes
is given by~\cite{Bassetto:1983mvz}
\begin{equation}
M^{{\lambda_k}}(\dots,p_i,p_k,p_j,\dots) = g_s
\left( \frac{p_i{\cdot}\epsilon_{\lambda_k}^*(p_k)}{p_i{\cdot}p_k} -
\frac{p_j{\cdot}\epsilon_{\lambda_k}^*(p_k)}{p_j{\cdot}p_k} \right)
M(\dots,p_\itilde,p_\jtilde,\dots)\,.
\label{eq:tech-soft-me}
\end{equation}
As in \cref{eq:collinear-factorization}, helicity indices that are not shown are the
same on both sides, and in particular with regards to the
$\itilde, i, \jtilde, j$ partons, there is an implicit constraint
$\delta_{\lambda_\itilde \lambda_i}\delta_{\lambda_\jtilde
  \lambda_j}$.
After expressing the polarisation vectors in terms of spinor products,
the soft matrix element in \cref{eq:tech-soft-me} can be written
as~\cite{Berends:1987me}
\begin{equation}
M^{{\lambda_k}}(\dots,p_i,p_k,p_j,\dots) =
\sqrt{2}g_s \frac{S_{-\lambda_k}(p_i,p_j)}{S_{-\lambda_k}(p_i,p_k)
S_{-\lambda_k}(p_j,p_k)}  M(\dots,p_\itilde,p_\jtilde,\dots) \,,
\label{eq:tech-soft-me-ii}
\end{equation}
again with the implicit
$\delta_{\lambda_\itilde \lambda_i}\delta_{\lambda_\jtilde \lambda_j}$
constraint.
Using the relations 
\begin{subequations}
  \begin{align}
    &S_{\lambda}(p, q) = - S_{\lambda}(q, p) = -S_{-\lambda}(p,q)^*, \\
    &|S_{\lambda}(p,q)|^2 = S_{\lambda}(p,q) S_{-\lambda}(q,p) = 2 \, p{\cdot}q,
  \end{align}
\end{subequations}
and additionally $S_{\lambda}(p_j, p_k) = \sqrt{(1-z)/z} S_{\lambda}(p_j,
p_i)$
when $i$ and $k$ are collinear, it is straightforward to show
that \cref{eq:tech-soft-me-ii} reduces to the soft-collinear
limit of \cref{eq:collinear-factorization,eq:branch-amp-spinor-prod},
noting that only contributions with $\lambda_\itilde=\lambda_i$
survive in the soft ($z\to 1$) limit of
\cref{eq:branch-amp-spinor-prod}.

For the application of the Collins--Knowles algorithm, the critical
element is the factorisation structure of the helicity indices in
\cref{eq:collinear-factorization}.
The observation that we make here is that it is possible to obtain the
correct soft arbitrary-angle limit for the amplitude without modifying
the factorisation of the helicity indices in \cref{eq:collinear-factorization}.
All that is needed to achieve this is to modify the spinor factors on
the right-hand side of \cref{eq:branch-amp-spinor-prod} so that
the $\itilde \to ik$ branching acquires a dependence on the
kinematics of $\itilde$'s dipole colour-partner $j$, without having to
introduce any dependence on the helicity of $j$.
We achieve this by replacing \cref{eq:branch-amp-spinor-prod}, in
the case where $\lambda_\itilde = \lambda_i$ with
\begin{equation}
  \label{eq:branch-amp-spinor-prod-soft}
  \mathcal{M}^{\lambda\, \lambda\,  \lambda_k}_{\tilde{\imath} \rightarrow i k} =
  \sqrt{2}g_s
  \mathcal{F}_{\itilde \to ik}^{\lambda\, \lambda\, \lambda_k}(z)
  \sqrt{\frac{1-z}{z}}
  \frac{S_{-\lambda_k}(p_i,p_j)}{S_{-\lambda_k}(p_i,p_k) S_{-\lambda_k}(p_j,p_k)}\,.
\end{equation}
One can also think of this as a correction factor to
  \cref{eq:branch-amp-spinor-prod}, computed as the ratio of
  \cref{eq:tech-soft-me-ii} and its collinear limit.
For the amplitude where $\lambda_\itilde \neq \lambda_i$ and for
$g \to q\bar q$ splittings, neither of which have a soft enhancement,
we retain \cref{eq:branch-amp-spinor-prod} as it is.
It is straightforward to verify that \cref{eq:branch-amp-spinor-prod-soft} reduces to
\cref{eq:branch-amp-spinor-prod} in the limit where $i$ and $k$
are collinear, and to \cref{eq:tech-soft-me-ii} when $k$ is
soft, i.e.\ $1-z \ll 1$.
When $k$ is soft and emitted from an $\itilde \jtilde$ dipole it is
somewhat arbitrary whether to account for the emission of $k$ as an
$\itilde \to ik$ branching in the spin-correlation tree or as a
$\jtilde \to jk$ branching.
That choice only makes a difference to the spin-correlation structure
for contributions with $\lambda_\itilde \neq \lambda_i$ or
$\lambda_\jtilde \neq \lambda_j$, which are both suppressed in the
limit $z \to 1$, or simply zero.

The numerical evaluation of the spinor products that appear in
\cref{eq:branch-amp-spinor-prod-soft} proceeds in the same fashion as was described in
Appendix A of Ref.~\cite{Karlberg:2021kwr}.
There, it was highlighted that the evaluation necessarily involves the
choice of a reference spinor direction, which causes the purely
collinear amplitudes to fail to reproduce the soft limit in an
azimuthally-dependent way.
\cref{eq:branch-amp-spinor-prod-soft} ensures that in the soft limit, the dependence on
that reference direction is eliminated.
We comment on this issue further in \cref{sec:app:reference-direction}.

A final observation is that the identification of the colour partner
$j$ is only unambiguous in the large-$\nc$ limit.
Accordingly, the algorithm as presented here only provides the correct
soft (non-collinear) spin-correlation structures within that
large-$\nc$ limit.
Recall that at large angles, the PanScales showers are anyway only
correct for the single (NLL) logarithms in the leading-$\nc$ limit,
even if residual subleading-$\nc$ corrections have been found to be
numerically small after application of the NODS colour
algorithm of Ref.~\cite{Hamilton:2020rcu}.
In \cref{sec:app:FC}, we will further comment on the size of residual
subleading-$\nc$ contributions for spin correlations.

\section{Fixed-order tests}
\label{sec:results}
\label{sec:fo-validation}

In order to demonstrate the effect of soft spin correlations, as well
as the correctness of our implementation, we will show below results
from the PanScales family of parton showers, at fixed order (this
section) and at all orders (\cref{sec:new-observable}).
As a reminder to the reader, the PanLocal shower uses a local recoil
scheme, where the emission of $p_k$ from a dipole
$\lbrace \tilde{p}_i, \tilde{p}_j \rbrace$ is associated with the
following momentum mapping,
\begin{subequations}
  \begin{align}  \label{eq:panlocal-map}
    p_k &= a_k \tilde{p}_i + b_k \tilde{p}_j + k_{\perp}\,, \\
    p_i &= a_i \tilde{p}_i + b_i \tilde{p}_j - f k_{\perp}\,, \\
    p_j &= a_j \tilde{p}_j + b_j \tilde{p}_j - (1-f) k_{\perp}\,,
  \end{align}
\end{subequations}
with $k_{\perp} = k_{\perp,1} \cos(\varphi) + k_{\perp,2}\sin(\varphi)$ and $-k_{\perp}^2 = k_t^2$.
The Sudakov components of the emission $p_k$ in
\cref{eq:panlocal-map} can be expressed in terms of the evolution
variable $v$ and an effective rapidity $\bar
\eta$,
\begin{equation}
\label{eq:panscales-variable-defs}
v = \frac{k_t}{\rho} e^{-\beta |\bar{\eta}|},
\qquad
\rho = \left(\frac{s_{\itilde}
    s_{\jtilde}}{Q^2 s_{\itilde\jtilde}}\right)^{\frac{\beta}{2}}\,,
\qquad
  a_{k} = \sqrt{\frac{s_{\jtilde}}{s_{\itilde\jtilde}s_{\itilde}}}\,
  k_t{e}^{+\bar \eta}\,,
  \qquad
  b_{k} = \sqrt{\frac{s_{\itilde}}{s_{\itilde\jtilde}s_{\jtilde}}}\,
  k_t{e}^{-\bar \eta}\,,
\end{equation}
where $s_{\itilde \jtilde} =2\ptilde_{i}\cdot \ptilde_{j}$,
$s_{\itilde}=2\ptilde_{i} \cdot Q$, the total momentum of the event is $Q$, and
the fixed parameter $\beta$ defines the angular scaling of $v$.
The other components are then fixed by requiring momentum conservation and on-shellness.
The PanLocal shower comes in a dipole variant, where $f = 1$, and in an antenna
variant, where
\begin{equation}
f = f(\bar{\eta}) = \frac{e^{2\bar{\eta}}}{1+e^{2\bar{\eta}}}\,.
\end{equation}
The PanGlobal shower instead uses the map
\begin{subequations}
  \begin{align}
    \bar p_k &= a_k \tilde{p}_i + b_k \tilde{p}_j + k_{\perp}\,, \\
    \bar p_i &= (1-a_k) \tilde{p}_i\,, \\
    \bar p_j &= (1-b_k) \tilde{p}_j\,,
  \end{align}
\end{subequations}
and applies a rescaling and Lorentz boost of all particles, including
$\bar{p}_{i,j,k}$, to restore four-momentum conservation.

As part of a validation of the implementation presented above, we compare the
effective differential cross section generated by the parton shower,
$\text{d}\sigma_{\text{PS}}$, to the soft, leading-colour cross section, $\text{d}\sigma_{\text{Exact}}^{\text{(LC)}}$, 
evaluated by considering the exact matrix element in those limits at fixed order in the strong
coupling $\mathcal{O}(\alpha_s^2)$ and $\mathcal{O}(\alpha_s^3)$.
The matrix
elements were computed analytically, with help from
FeynCalc~\cite{MERTIG1991345,Shtabovenko:2016sxi,Shtabovenko:2020gxv} and 
MultivariateApart~\cite{Heller:2021qkz}, using
\textsc{Mathematica}~\cite{Mathematica}, in the limit where the emitted
particles are much softer than the original $q\bar q$ pair. They were
checked against matrix elements obtained using amplitudes from
Refs.~\cite{Nagy:1998bb,Badger:2005jv} and were found to agree at permille
level in the quasi-soft limit.\footnote{Specifically $z_{1} =
  10^{-4}$ and $z_{2} = 10^{-7}$, with the residual permille difference
  relative to the full matrix elements being consistent with
  corrections of order $z_1$ and $z_2/z_1$.
  The separate soft-limit matrix elements facilitate tests 
  at asymptotic values of the softness.
}
In principle, one could also extract the relevant limits from
Refs.~\cite{Catani:1999ss}.
The squared matrix elements are given for completeness in \cref{app:analytic-mes}.
In the following, all results are shown in the large-$\nc$ limit, which is
achieved in the parton shower by setting $C_F = \frac{1}{2} C_A =
\frac32$.
We comment on the magnitude of subleading-colour effects in
\cref{sec:app:FC}, where the full-colour cross sections are
compared with results obtained by the PanScales methods set out in
Ref.~\cite{Hamilton:2020rcu} for including a subset of
subleading-colour corrections in dipole showers.

\begin{figure}
\centering
\includegraphics{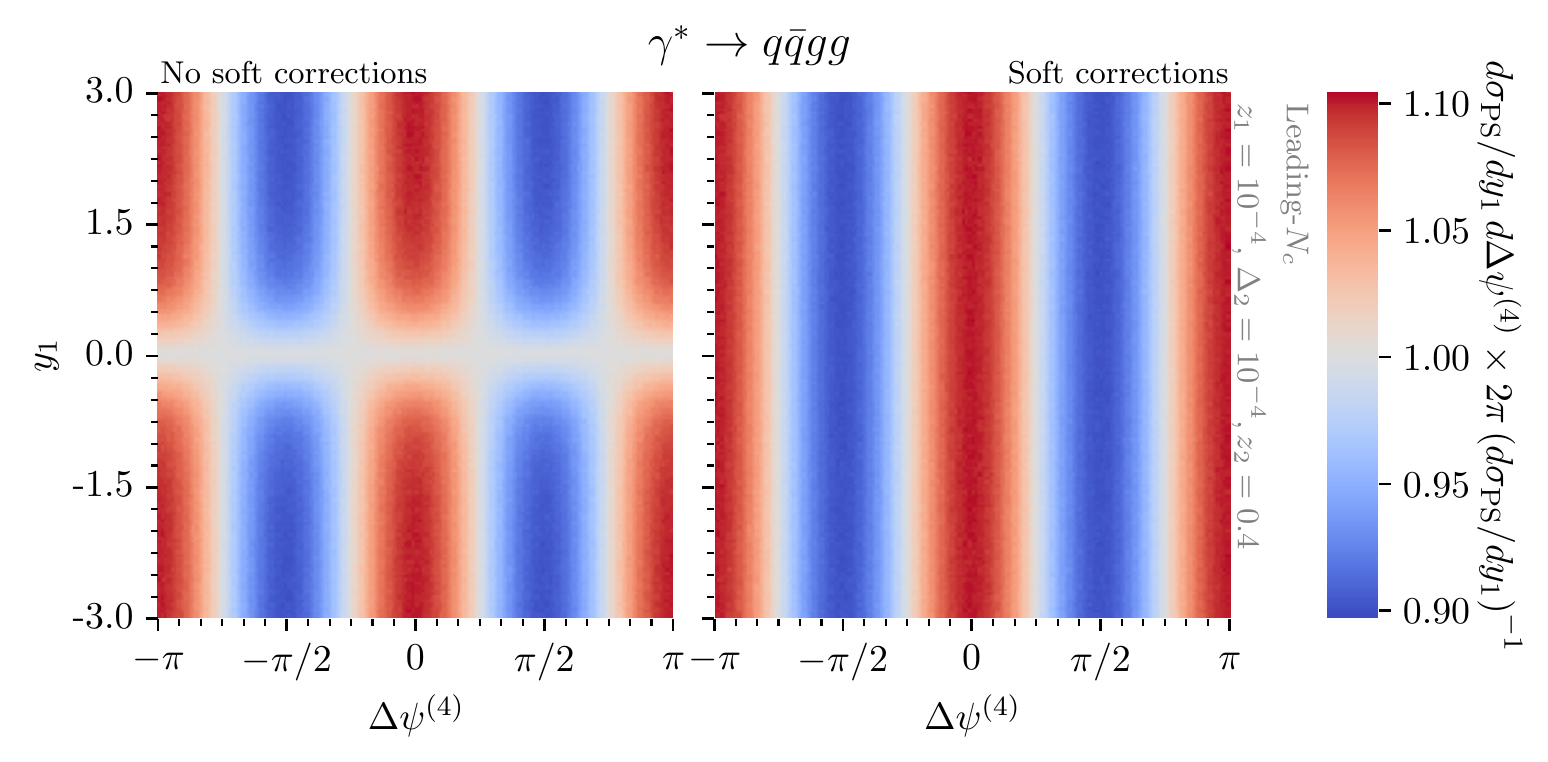}
\includegraphics{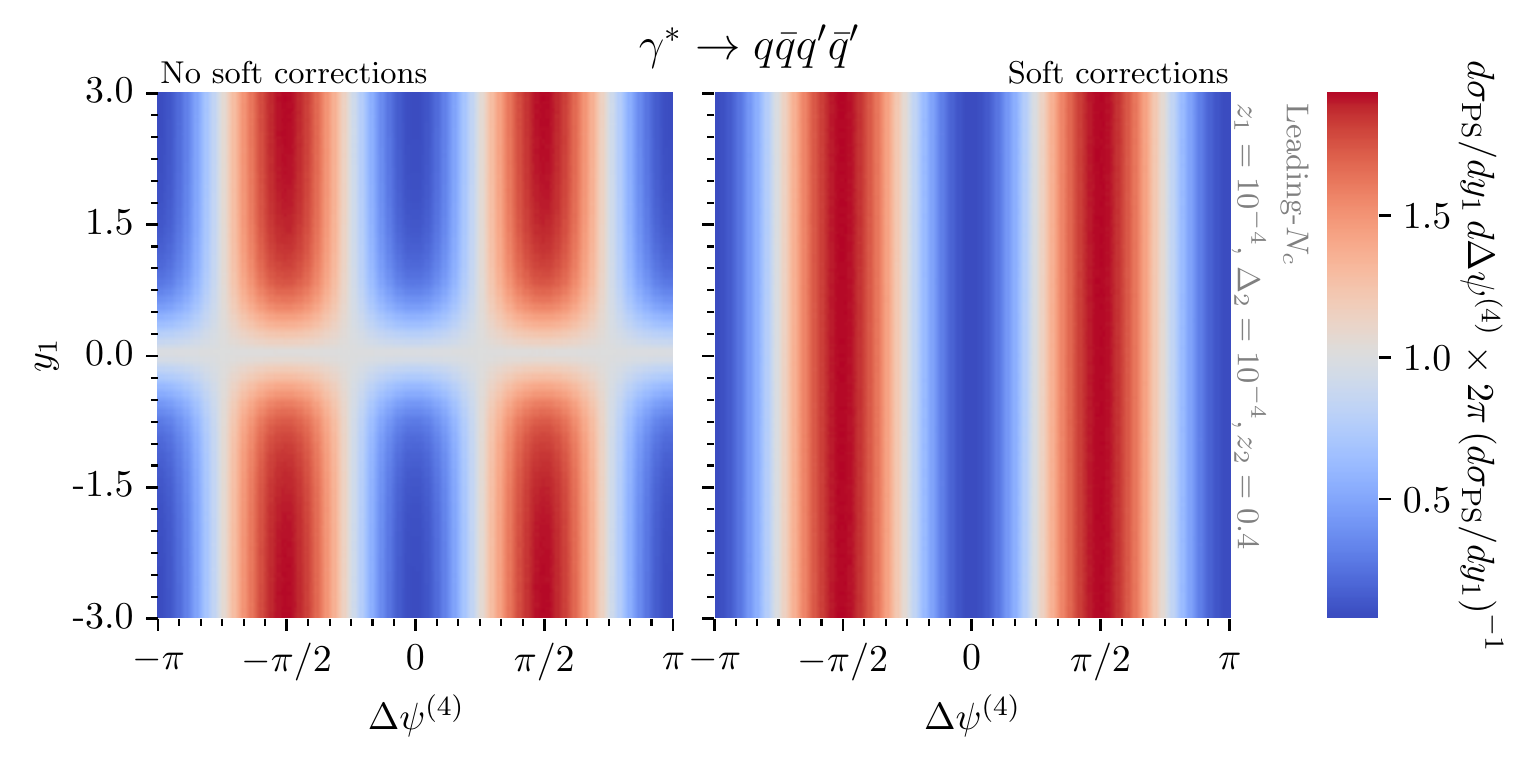}
\caption{ The double-differential cross section
  $d\sigma_{\text{PS}}/dy_1 d\Delta\psi^{(4)}$ for $e^+e^- \to
  q\bar q gg$ (top) and $e^+e^- \to q\bar q q' \bar q'$ (bottom),
  without soft spin corrections (left) --- i.e. purely collinear spin
  correlations --- and with soft corrections enabled (right),
  normalised to the single-differential cross section
  $d\sigma_{\text{PS}}/dy_1$. Comparisons are carried out in the
  leading-$\nc$ approximation with $C_F=\frac{C_A}{2}=\frac{3}{2}$.}
\label{fig:results-alphas2}
\end{figure}
\begin{figure}
\centering
\includegraphics[page=2]{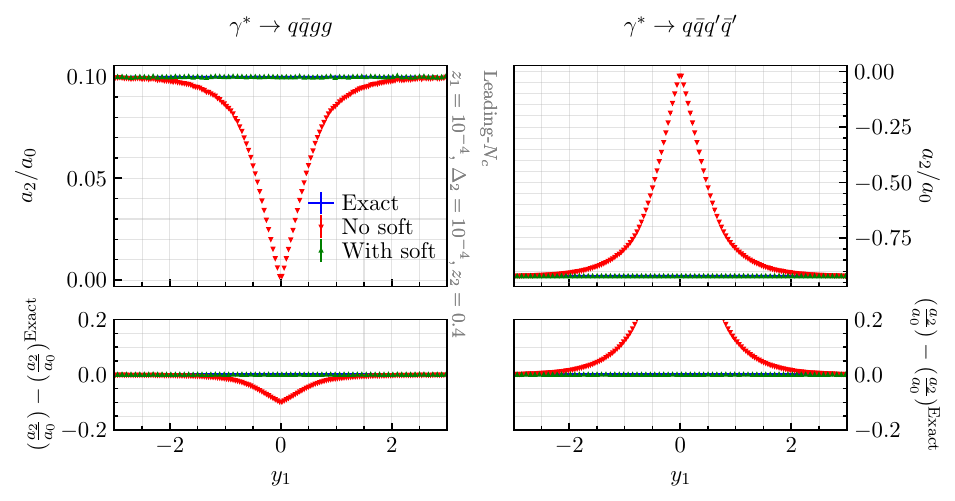}
\caption{The extracted ratio $a_2/a_0$, as defined in \cref{eq:a0a2-fixed-order}, 
from the double-differential cross section of \cref{fig:results-alphas2}, as well as the 
exact result, as a function of $y_1$. Comparisons are carried out in the
  leading-$\nc$ approximation with $C_F=\frac{C_A}{2}=\frac{3}{2}$.
}
\label{fig:results-alphas2-FT}
\end{figure}

First, we examine configurations at second order, like that shown in
the upper-left part of 
\cref{fig:results-alphas-conf}, where a soft, wide-angle gluon
$g_1$ is emitted from the original $q \bar q$ dipole,
and then splits collinearly, either as $g_1 \to gg$ or $g_1 \to q' \bar
q'$.
We fix the energy fraction carried away by the gluon $g_1$ in the
first splitting, $z_1 = E_1/Q = 10^{-4}$, as well as the collinear
momentum fraction and splitting angle of the second collinear
branching, $z_2 = 0.4$ and $\Delta_2 = 10^{-4}$ respectively. We then
sample over the rapidity $y_1$ of the gluon $g_1$, and over the
azimuthal angles of both splittings.
\cref{fig:results-alphas2} shows the differential cross section, as
generated by the shower, as a function of the rapidity
$y_1$ of the emitted gluon (vertical axis) and $\Delta\psi^{(4)}$
(horizontal axis), which corresponds to the reconstructed difference
in the azimuthal angles of the primary $q \to qg_1$ and secondary
$g_1 \to gg$, or $g_1 \to q'\bar q'$ splitting planes (see also the
right panel of Fig.~\ref{fig:results-alphas-conf}).
The upper panels are for the $\gamma^* \to q\bar q gg$ process, the
lower panels are for $\gamma^* \to q\bar q q'\bar q'$.
The left hand panels use just the collinear spin correlations of our
earlier work~\cite{Karlberg:2021kwr}, while the right-hand panels show
the results when we include the soft spin correlations of this work.
The striking difference between left and right-hand panels is that,
with the soft spin correlation corrections, the azimuthal modulation
is independent of $y_1$, while with the pure collinear
implementation of the spin correlations, that is not the case, with
spurious structure appearing in the large-angle region where the
shower switches between $g_1$ being emitted by the $q$ dipole end to
it being emitted from the $\bar q$ dipole end.
The independence on $y_1$ in the soft case is the correct behaviour.
This is easy to understand intuitively: soft gluon emission and subsequent
splitting should be invariant under longitudinal boosts along the
parent ($q\bar q$) dipole direction.

To demonstrate that the result is indeed correct not just in structure but
also normalisation, we take the Fourier cosine transform with respect
to $\Delta \psi^{(4)}$, i.e.\ we extract the values of $a_0(y_1)$ and $a_2(y_1)$, 
following 
\begin{equation}
  \label{eq:a0a2-fixed-order}
  \frac{d\sigma}{dy_1 d\Delta\psi^{(4)}}
  = a_0(y_1) + a_2(y_1) \cdot \cos \big( 2\Delta\psi^{(4)} \big)\,,
\end{equation}
which encodes the correct azimuthal structure of spin-dependent observables~\cite{Karlberg:2021kwr}.
\cref{fig:results-alphas2-FT} shows the ratio of $a_2/a_0$ for
$\gamma^* \to q\bar q gg$ (left) and $\gamma^* \to q\bar q q'\bar q'$
(right) as a function of $y_1$, comparing to the exact result.
There we see the excellent agreement of our soft-spin correlation
procedure with that matrix element.
Note the usual features that $g_1 \to q'\bar q'$ splittings peak in a
plane that is perpendicular to the $q\bar q g_1$ plane ($a_2<0$), and with a
substantially stronger modulation than the $g_1 \to gg$ splitting,
which peaks in the same plane as the $q\bar q g_1$ ($a_2 >
0$).

\begin{figure}
\centering
\includegraphics{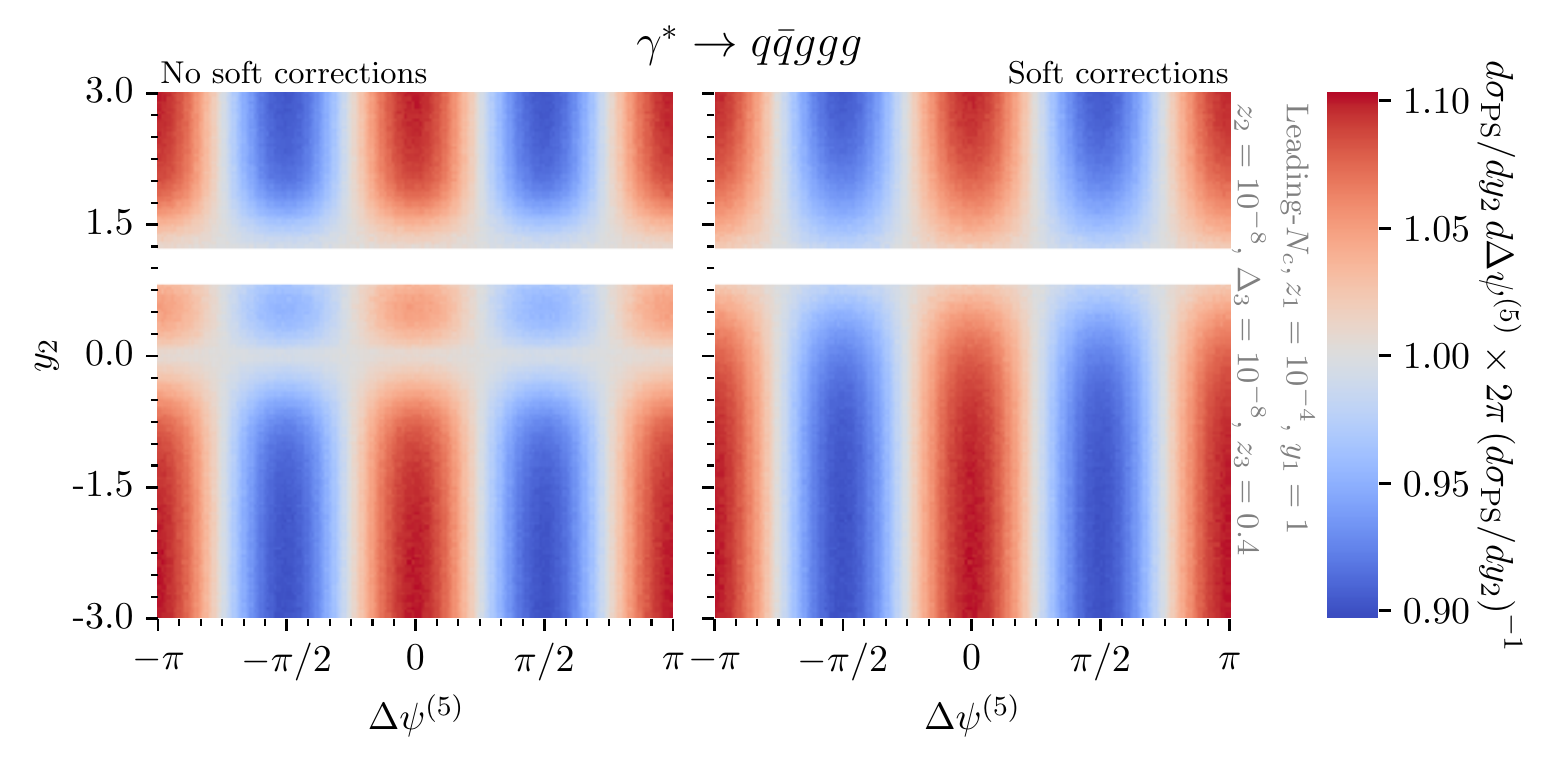}
\includegraphics{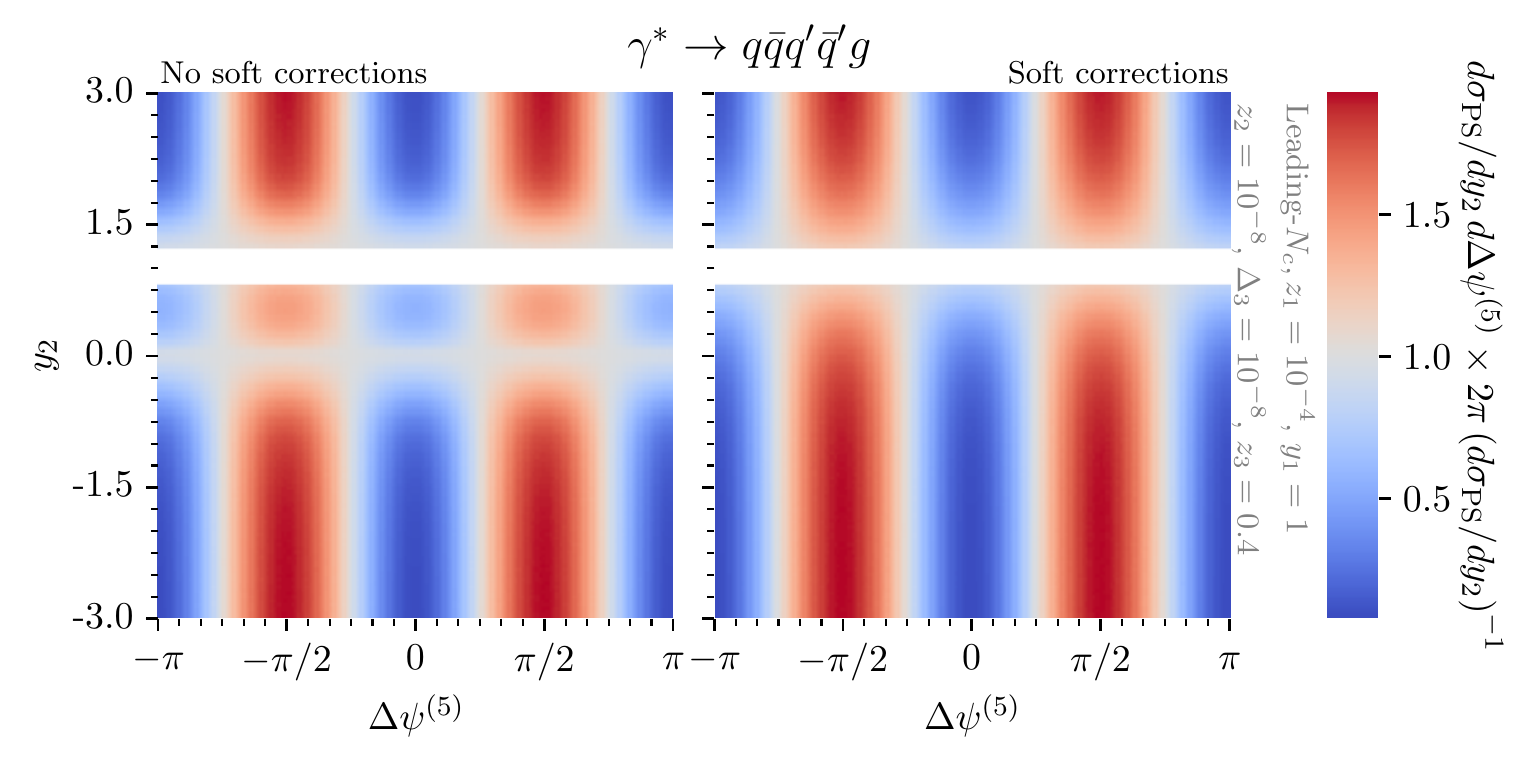}
\caption{Same as \cref{fig:results-alphas2} for $e^+e^- \to q\bar q
  ggg$ (top) and $e^+e^- \to q\bar q q' \bar q' g$ (bottom), as a
  function of the rapidity $y_2$ of a soft gluon --- which then
  splits collinearly --- and the azimuthal angle $\Delta \psi^{(5)}$
  between the primary ($q \to q g_2$) and secondary ($g_2 \to gg$ or
  $g_2 \to q' \bar q'$) splittings. Comparisons are carried out in the
  leading-$\nc$ approximation with $C_F=\frac{C_A}{2}=\frac{3}{2}$.}
\label{fig:results-alphas3}
\end{figure}
\begin{figure}
\centering
\includegraphics[page=2]{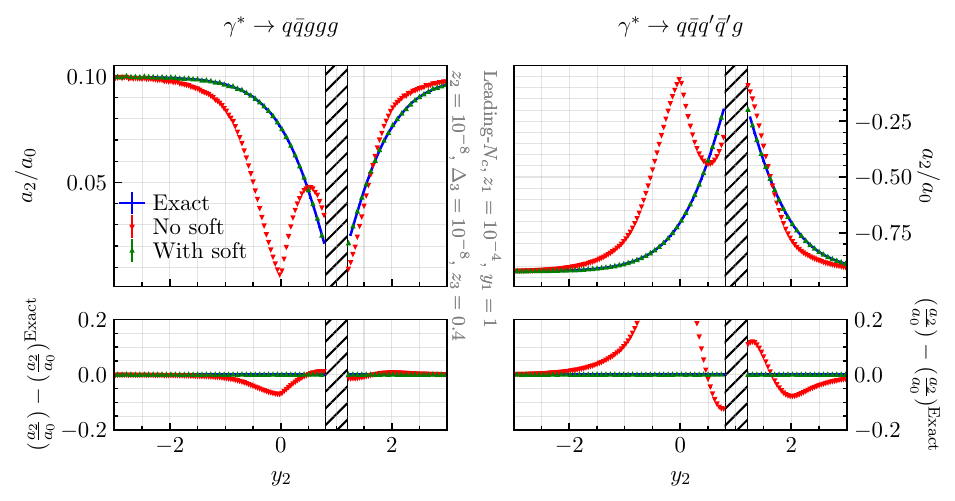}
\caption{Same as \cref{fig:results-alphas2-FT}, but for the configurations used in \cref{fig:results-alphas3}.}
\label{fig:results-alphas3-FT}
\end{figure}

In \cref{fig:results-alphas3,fig:results-alphas3-FT},
we repeat the tests for 5-parton processes, $\gamma^* \to q\bar q gg g_1$
and $\gamma^* \to q\bar q q'\bar q'g_1$,
cf.\ \cref{fig:results-alphas-conf} (lower left).
We fix the rapidity of the first (non-splitting) gluon $y_1=1$ and
its energy fraction $z_1 = 10^{-4}$ and integrate over its azimuth.
The second gluon, with energy fraction $z_2 = 10^{-8}$, splits to a
collinear $gg$ or $q'\bar q'$ with opening angle $\Delta_3 =
10^{-8}$.
\cref{fig:results-alphas3} shows the result of sampling over its
rapidity $y_2$ (vertical axis) and the azimuthal angular
difference between its decay plane and the $q\bar q g_2$ plane,
$\Delta \psi^{(5)}$.
We mask a small strip around $y_2 \sim y_1$, a region that is
affected by a divergence when $g_1$ and $g_2$ become collinear (this
region is correctly treated by the soft spin procedure, but
would be more effectively probed by a separate analysis).
The corresponding results for $a_2/a_0$ are shown in
\cref{fig:results-alphas3-FT}.
The exact matrix element has non-trivial structure for $y_2$ close
to $y_1$, which is well reproduced by our soft-spin procedure.
The purely collinear spin procedure introduces spurious structures
for $y_2\simeq0$ and $y_2 \simeq y_1$, corresponding to the
transition regions between different dipole ends.

As mentioned above, all the results shown in this section (and in the
next section) are in a large-$\nc$ approximation whereby $C_F = C_A/2 =
3/2$.
Subleading-$\nc$ corrections affect not just the intensity of soft
gluon emission, but also the spin structure.
The reason is that each dipole that contributes to radiating the soft
gluon can transmit distinct spin information to that soft
gluon.
The impact of this is illustrated at fixed order in
\cref{sec:app:FC}, where we see that residual full-$\nc$
effects are numerically small, of the order of $\lesssim 1\%$ ($3\%$) for the
$gg$ ($q\bar q$) channel with the NODS procedure.

\section{A new observable sensitive to soft spin effects}
\label{sec:new-observable}

Soft spin correlations are an essential requirement within the
PanScales NLL conditions, and we expect that machine-learning
approaches to jet analyses and substructure are likely to be learning
some features of the azimuthal correlations that they induce.
However, we are not aware of any existing observables that are
directly sensitive to these effects.
Accordingly, in this section, we propose an observable that probes
both soft wide-angle and subsequent collinear splittings, and use the
asymptotic limit of our parton showers to provide a reference
single-logarithmic resummation for its distribution.

\subsection{Definition of the observable}

We start by clustering an $e^+e^-$ event,
in its centre-of-mass frame, using the spherical version of the
Cambridge/Aachen (C/A)
algorithm~\cite{Dokshitzer:1997in,Wobisch:1998wt} with $E$-scheme
($4$-momentum) recombination, as implemented in
FastJet~\cite{Cacciari:2011ma}.
We undo the steps of the clustering sequence so as to obtain exactly
two jets, with momenta $p_1$ and $p_2$ and use $\vec p_1 - \vec p_2$
to define the event axis direction $\hat a$.
For each particle or jet $i$, we identify its rapidity with respect to the
event direction by evaluating the $3$-momentum component parallel to
$\hat a$, $p_{i,\hat a}$ and using the usual definition $y_i = \frac12 \ln \left( (E_i
+ p_{i,\hat a})/(E_i - p_{i,\hat a})\right)$.
We will be interested in particular in subjets that are within a slice
$|y| < y_{\max}$ where $y_{\max}$ is a parameter of order $1$.

\begin{figure}
\centering
\includegraphics[width=.55\textwidth]{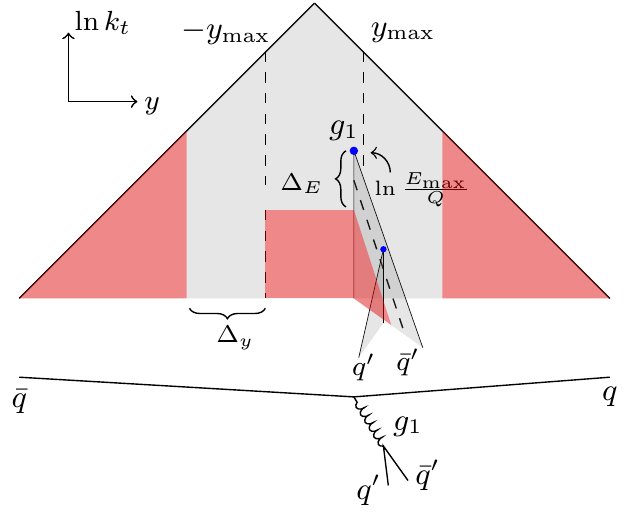}
\caption{For the definition of the slice observable $\dpsisl$, we consider
  primary splittings $q\to qg_1$ whose softer branch is found in a central
  rapidity slice $[-y_{\text{max}},y_{\text{max}}]$. Following that softer
  branch, we select secondary splittings that satisfy $z \geq \zcut$. The cuts are
  indicated by dashed lines on the Lund plane. The red shaded patches illustrate the 
  regions where shower emissions may be vetoed in the limit of $\alpha_s \rightarrow 0$
  at a fixed $\lambda=\alpha_s L$.
}
\label{fig:lund-slice}
\end{figure}

We then carry out a Lund-diagram~\cite{Andersson:1988gp} style
analysis of the event using the $e^+e^-$ C/A declustering tree, in the
spirit of Ref.~\cite{Dreyer:2018nbf}.
The observable is defined through the following steps
\begin{enumerate}
\item We examine all C/A declusterings and identify any that satisfy
  the property that the harder subjet $j$ has $|y_j| > y_{\max}$ and
  the softer subjet $k$ has $|y_k| < y_{\max}$ (softer and harder are
  defined in terms of the magnitudes of the $3$-momenta).
  If there is no such declustering, the event does not contribute to the
  final histogram.
  If there is more than one declustering that satisfies that property,
  we choose the one with largest $k_t = |\vec p_k| \sin
  \theta_{jk}$.\footnote{One might also choose to define 
  $k_t =|\vec p_k| \sqrt{2(1-\cos\theta)}$, which is a monotonic function.}
  We denote the $k_t$ associated with this declustering as $k_{t,1}$
  and the azimuth as $\psi_1$ (azimuths are defined following the
  procedure in Ref.~\cite{Karlberg:2021kwr}).

\item Next we consider the declustering tree of $k$, and specifically
  all declusterings that belong to $k$'s Lund leaf (i.e.\ at each
  declustering, the softer subjet belongs to the leaf, and we then
  recursively continue to identify other Lund leaf subjets of $k$ by
  following the further declustering of the harder subjet).
  For each $\ell \to mn$ declustering, we evaluate
  $z = |\vec p_n|/(|\vec p_m| +|\vec p_n|)$ where $n$ is the softer
  branch.
  We identify all declusterings with $z$ larger than some parameter
  $z_\text{cut}$.
  If there are none, the event does not contribute to the
  final histogram.
  If there is more than one, we choose the one with largest $k_t$,
  which we denote as $k_{t,2}$.
  We denote its azimuth as $\psi_2$.
  \label{obs:step2}
\item Finally, for all events with $k_{t,2} > k_{t,\text{min}}$, we bin
  the distribution of the signed azimuthal angle
  $\dpsisl \equiv \psi_2 - \psi_1$ between the two splitting planes
  identified above.
\end{enumerate}
An illustration of how this observable works at the first non-zero
order, $\as^2$, is given in \cref{fig:lund-slice}, which shows a
primary emission $g_1$ within the $y_{\max}$ rapidity constraint and
its subsequent collinear splitting to a $q'\bar q'$ pair with
$z > z_\text{cut}$.
In this simple case, $\dpsisl$ is the difference in azimuthal angle
between the $g_1q$ plane and the $q'\bar q'$ plane.
Code to evaluate the observable will be distributed as part of the
LundPlane FastJet contrib package.

At order $\mathcal{O}(\alpha_s^2)$, the leading-$\nc$ contribution to the slice observable is given
by
\begin{multline}
\frac{d\sigma^\text{FO}}{d\dpsisl} = |M_{q\bar q}|^2 \,
2y_\text{max} \left(\frac{\alpha_s}{\pi}\right)^2
C_F\,\text{ln}^2\left(\frac{k_{t,\text{min}}}{Q}\right) \\ \times
\int_{z_\text{cut}}^{1-z_\text{cut}} \text{d}z_2 \mathcal{P}_{kg}(z_2)\big[ 1 +
B_{kg}(z_2) \cos \left(2 \dpsisl \right) \big] \,,
\label{eq:fixed-order-slice-1}
\end{multline}
with $\mathcal{P}_{kg}(z)$ the usual leading-order unregularised DGLAP splitting
functions, with their symmetry factors, where $k \in \lbrace g, q \rbrace$ refers to
the secondary splitting, $g_1 \to
gg$ or $g_1 \to q\bar q$, and the $B_{kg}(z)$ as given in Ref.~\cite{Karlberg:2021kwr},
\begin{align}
\mathcal{P}_{gg}(z) &= C_A \left(\frac{z}{1-z} + \frac{1-z}{z} + z(1-z) \right), &B_{gg}(z) &= \frac{z^2 (1-z)^2}{(1-z(1-z))^2}\,, \\
\mathcal{P}_{qg}(z) &= T_R n_f (z^2 + (1-z)^2), &B_{qg}(z) &= \frac{-2z(1-z)}{1-2z(1-z)}\,.
\label{eq:fixed-order-slice-2}
\end{align}
The relative azimuthal modulations are maximal for $z=1/2$, and therefore adjusting
$z_\text{cut}$ in \cref{eq:fixed-order-slice-1} affects their ultimate observed size.

At all orders, the dominant logarithmically enhanced terms for this
observable are single-logarithmic contributions $\as^n L^n$ with
$L = \ln k_{t,\min}/Q$.
The single-logarithmic resummation for the observable has the
structure
\begin{equation}
  \label{eq:all-order-general-form}
  \frac{1}{\sigma} \frac{d\sigma}{d\dpsisl} =
  a_0(\as L) + a_2(\as L) \cdot \cos \left(2\dpsisl\right) + \order{\as^n L^{n-1}}\,.
\end{equation}
The resummation for the analogous, purely collinear $\Delta \psi_{12}$
of Ref.~\cite{Karlberg:2021kwr} could be obtained using a collinear
spin extension of the MicroJets
code~\cite{Dasgupta:2014yra,Dasgupta:2016bnd}.
This then served to validate the PanScales shower at
single-logarithmic order.
The MicroJets code cannot, however, be used for soft spin effects.
Instead, for the validation of our implementation of soft-spin effects
in the PanScales showers, we rely on the fixed-order tests of
\cref{sec:fo-validation}.
Here we use the PanScales showers to determine a reference resummation
for the soft-spin observable.

\subsection{Strategy for resummation with the PanScales showers}

As in previous PanScales work, our strategy to obtain the resummed
result for the soft-spin observable is to consider the limit where we
take the strong coupling constant $\as \to 0$, with
$\as \ln k_{t,\min} / Q$ held fixed.
%
In practice, we use a fixed small value of the $\alpha_s(Q) = 10^{-7}$
with 1-loop running (with $n_f = 5$ light flavours), for different
values of $\lambda = \alpha_s L$.

Running the PanScales showers at such small values of $\alpha_s$ and large values of 
the logarithm is made possible by the techniques described in Appendix D of 
Ref.~\cite{Karlberg:2021kwr}. 
Additionally, in order to maintain computationally manageable
multiplicities, Ref.~\cite{Karlberg:2021kwr} ran the shower in a
version where soft emissions were vetoed.
This could be done safely, as the observables under consideration were
only sensitive to collinear emissions.
On the other hand, $\dpsisl$ is also sensitive to soft emissions in
the slice and removing all soft emissions would yield the incorrect
results.
Instead, therefore, we allow emissions to be generated if either they
are at (absolute) rapidity with respect to either dipole parent that
is below some threshold $|y| < \Delta_y + y_\text{max}$
($\Delta_y$ should be taken substantially larger than the rapidity
$y_\text{max}$ of the slice with respect to the event axis), or if
the parent is in the slice and the emission has an energy that is
above some threshold $\ln E/Q > \Delta_{E} + \ln E_\text{max}/Q $,
where $E_\text{max}$ is the energy of the highest-energy emission in
the slice.
This is illustrated in Fig.~\ref{fig:lund-slice}.
We have verified that if we work at finite values of $\as$ and
$\ln k_{t,\min}/Q$, results with and without these vetoes are
identical to within statistics.
\logbook{7de3c96d1}{See code in
  2020-eeshower/analyses/soft-spin-all-order/EmissionVetoSoftSpin.hh
  for the actual implementation.}

One further subtlety concerns the handling of flavour.
As we have seen at fixed order, soft spin correlation effects depend
strongly on the flavour structure of the ultimate collinear splitting
that we examine and it is of interest to present resummed results
separated according to the flavour structure. 
In Ref.~\cite{Karlberg:2021kwr}, to separate channels we kept track of
the flavour of emitted particles in the spin correlation tree itself.
Here we take a different approach: we identify the flavour of
pseudo-jets appearing in the Lund declustering sequence by summing the
flavours of the individual constituents.
This definition of flavour is infrared unsafe, with the infrared
divergence contributing at order $\as^n \ln^{n-1} k_{t,\text{cut}} / Q$
where $k_{t,\text{cut}}$ is the shower's infrared
cutoff.\footnote{Specifically, for any collinear splitting, one can
  dress that splitting with an additional much softer gluon that then
  splits to a $q\bar q$ pair at angles commensurate with the collinear
  splitting.
  The individual quarks in that soft $q\bar q$ pair can separately
  contaminate the net flavour of one or other of the collinear prongs.
  The likelihood of such a contamination is driven by the soft
  divergence of the gluon emission, which generates one logarithm.
  Since the $g \to q\bar q$ splitting has no further soft divergence and
  because all of the $g \to q\bar q$ angles are constrained to be of
  the same order as the original collinear splitting, there are no
  further logarithms.
  Therefore the overall structure of this divergence involves a factor
  $\as^2 \ln k_{t,\text{cut}} / Q$, which at higher orders can involve
  further powers of $\as \ln k_{t,\text{cut}} / Q$.
  This divergence is essentially the same as that discussed in the
  context of the flavour-$k_t$ algorithm~\cite{Banfi:2006hf}.
  One might consider using the flavour-$k_t$ algorithm
  together with Lund declustering, but caution is needed, because the $k_t$ family of
  algorithms~\cite{Catani:1991hj} generates double-logarithmic
  structures in the Lund plane~\cite{Dreyer:2018nbf}.  }
In most circumstances this prevents the use of flavour in conjunction
with the Cambridge/Aachen algorithm.
However, in our study here, we consider the limit $\as \to 0$ with
$\as \ln k_{t,\min} / Q$ held fixed.
Furthermore, in this limit we are free to choose
$k_{t,\text{cut}} \sim k_{t,\min}$.
This ensures that the formally divergent terms
$\as^n \ln^{n-1} k_{t,\text{cut}} / Q$ actually vanish in our $\as \to 0$
limit.
This is a feature that we can exploit only for studies at
single-logarithmic level.
Were we to examine further subleading logarithms, we would need to
identify an alternative approach to flavour.

Given the net flavour of the pseudo-jets appearing in the Lund
declustering sequence, we assign a secondary declustering to:
\begin{itemize}
\item the $gg$ channel, if both the harder and softer branches of the secondary
declustering are flavourless,
\item the $q\bar q$ channel, if each of the two prongs has the net
  flavour of a single quark or anti-quark, and the sum of those prongs
  has no net flavour,\footnote{In the case of a multi-flavoured
    pseudo-jet, the contribution is binned in the rest channel.
    The multi-flavoured contribution is suppressed by a
    relative power of $\as$.
  }
\item the \textit{rest} channel, otherwise (which mainly involves
  $q\to qg$ splittings).
\end{itemize}

Unless specified, the results that we will show are obtained with the
PanGlobal shower with $\beta = 0$.

\subsection{Resummed results}
\label{sec:resummed-result}

\begin{figure}
\centering
\includegraphics{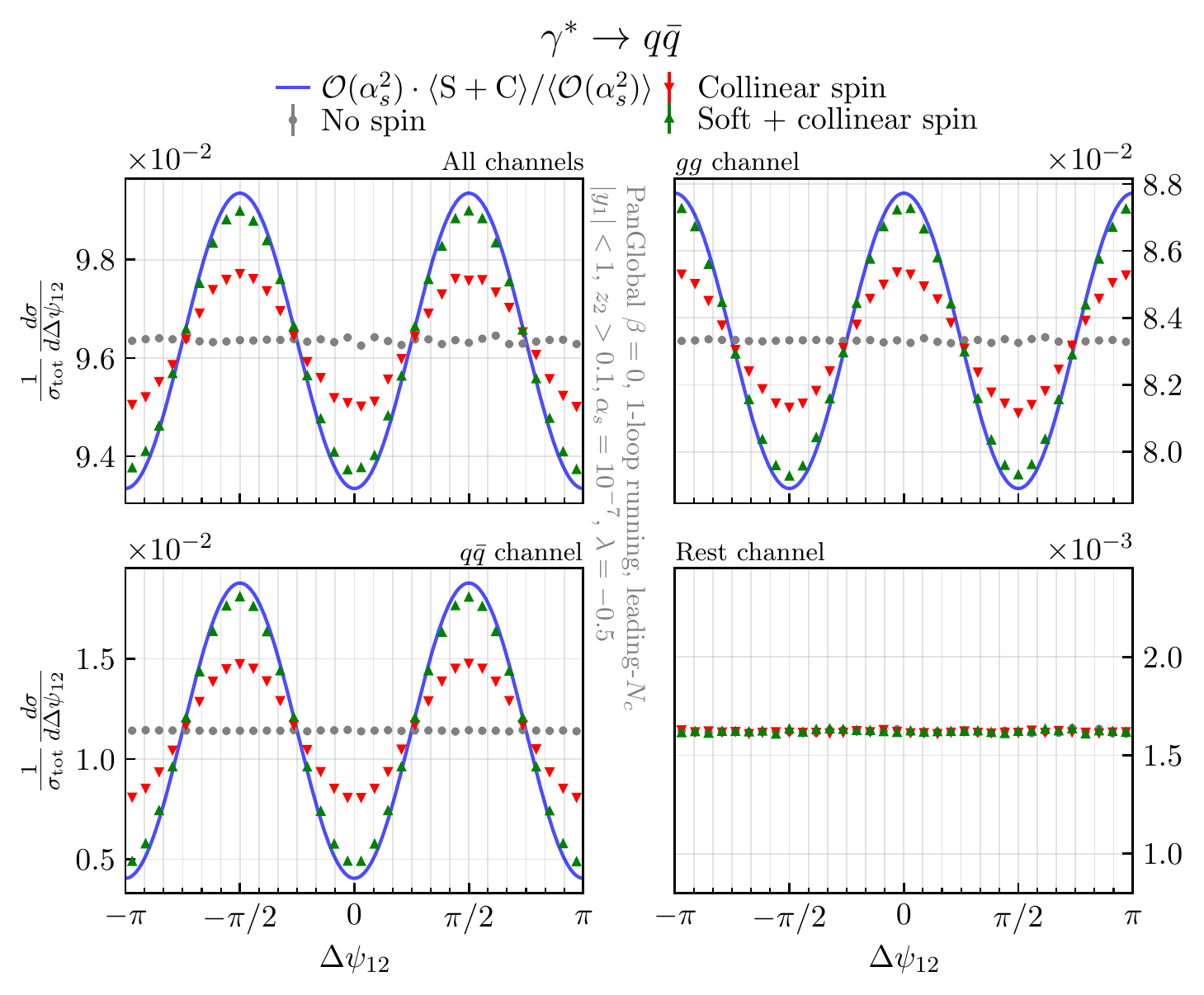}
\caption{All-order comparison for the $\dpsisl$ observable defined in
  the text for $\gamma^*\rightarrow q\bar{q}$. The four panels show
  the combination of all flavour channels for the splitting in the
  slice (upper left), the $g\rightarrow gg$ channel (upper right), the
  $g\rightarrow q\bar{q}$ channel (lower left), and a remainder
  channel where the splitting is classified neither as $gg$ nor
  $q\bar{q}$ (lower right).
  The plots show the predictions without spin correlations (grey dots), with
  the collinear spin correlations of Ref.~\cite{Karlberg:2021kwr}
  (red triangles), and with the soft gluon corrections of this work (green triangles).
  The blue curve is the $\mathcal{O}(\alpha_s^2)$ result, rescaled so that its
  mean value coincides with the all-order result.
  The results are produced with the PanGlobal shower, with $\beta =
  0$, in a leading-$\nc$ approximation
  with $C_F=\frac{C_A}{2}=\frac{3}{2}$, and using $\lambda \equiv \as L
  = -0.5$.
} 
\label{fig:allorder-dpsi}
\end{figure}

All-order results for the observable $\dpsisl$ are displayed in
\cref{fig:allorder-dpsi} for $\lambda =
\as \ln k_{t,\min} / Q =-0.5$, where the central rapidity
slice is given by $|y_1| < y_{\text{max}}=1$ and the cut on the energy
fraction of the secondary splitting is $z_2 > z_{\text{cut}} = 0.1$.
For the individual emission vetoes described above, we choose
$\Delta_y = 9$, $\Delta_{E} = -10$.
We use the leading-$\nc$ limit of $C_F = C_A/2 = 3/2$.
Results are shown for all flavour channels combined (top left), and
separately in the $g \to gg$ (top right), $g \to q\bar q$ (bottom
left) and ``rest'' channels (bottom right).
In the case where no spin correlations are applied (grey bullets), all curves
show a uniform distribution in $\dpsisl$.
If the purely-collinear spin correlations are enabled, but without
soft corrections (red triangles), we observe an azimuthal modulation of order
$\sim 1.4\%$ in the distribution where all splitting channels are
summed.
Once the soft corrections are enabled (green triangles), the spin correlations
have a distinctly larger effect, with a modulation of $\sim 2.8\%$ in the
all-channels distribution ($58\%$ in the $g \to q\bar q$ channel).
For comparison, we also show the distribution obtained at second order
(blue line), rescaled so that its mean value coincides with the mean
value of the all-order result.
It helps to illustrate that relative to that fixed-order result, the
resummation leads to a modest reduction in the degree of spin
correlations (it also affects the
overall normalisation, reducing it by 22\% compared to the fixed order).
Our interpretation of this observation is that, as in the purely
collinear case~\cite{Karlberg:2021kwr}, the spin correlations are
partly washed out by the resummation, as spin information is scattered
across the event by multiple gluon emissions along the declustering
sequence.

\begin{figure}
  \centering
  \includegraphics{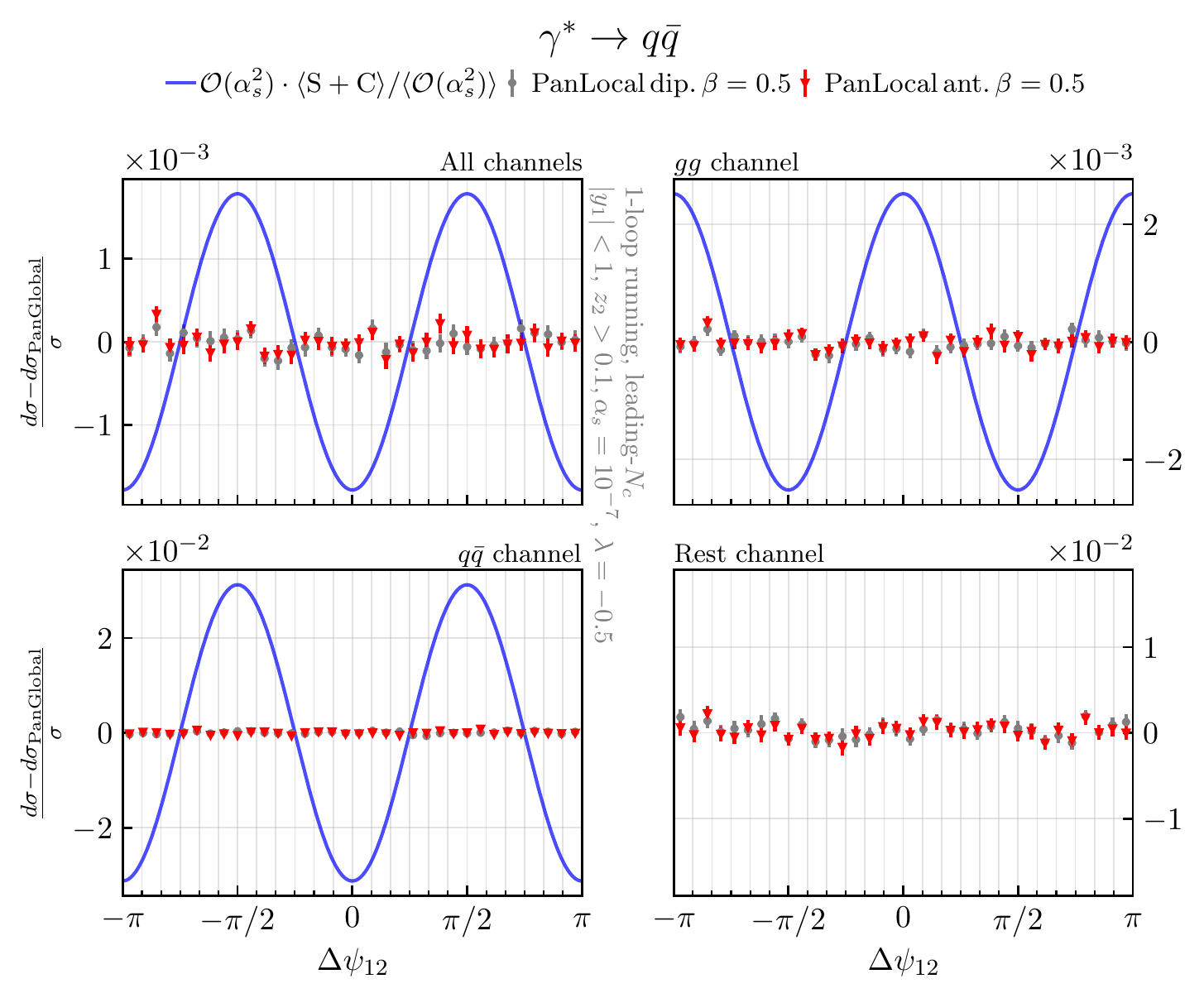}
  \caption{Same as Fig.~\ref{fig:allorder-dpsi}, but where the all-order 
    results with soft spin corrections are produced by two different 
    PanScales showers, PanLocal dipole with $\beta = 0.5$ (grey) 
    and PanLocal antenna with $\beta = 0.5$ (red). 
    Instead of the normalised cross section, we show the normalised difference with 
    respect to the PanGlobal $\beta = 0$ results of Fig.~\ref{fig:allorder-dpsi}.}
  \label{fig:allorder-dpsi-shower-comparison}
\end{figure}

In \cref{fig:allorder-dpsi-shower-comparison} we compare the
PanGlobal shower with $\beta=0$ to the dipole and antenna versions of
the PanLocal shower with $\beta=0.5$. Given that we do not know the
analytical resummation of the $\dpsisl$ observable, this is an
important cross-check that no subleading logarithmic effects are
present in our setup, and helps give us confidence that our
implementation reproduces the correct all-order structure.

\begin{figure}
  \centering
  \begin{subfigure}{0.55\textwidth}
    \centering
    \includegraphics[width=0.98\textwidth]{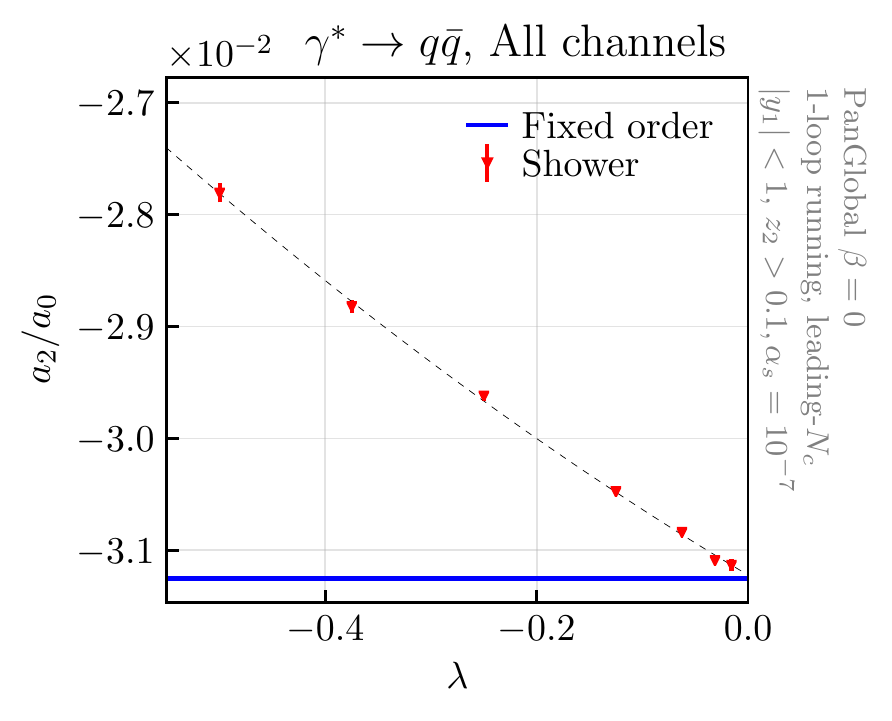}
  \end{subfigure}
  \begin{subfigure}{0.48\textwidth}
    \centering
    \includegraphics[width=1.\textwidth]{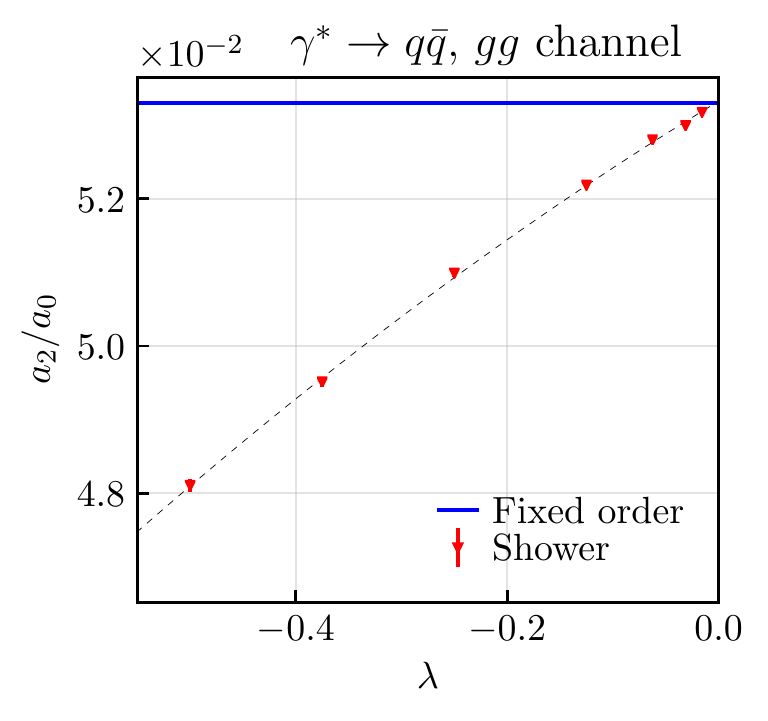}
  \end{subfigure}
  \begin{subfigure}{0.48\textwidth}
    \centering
    \includegraphics[width=1.\textwidth]{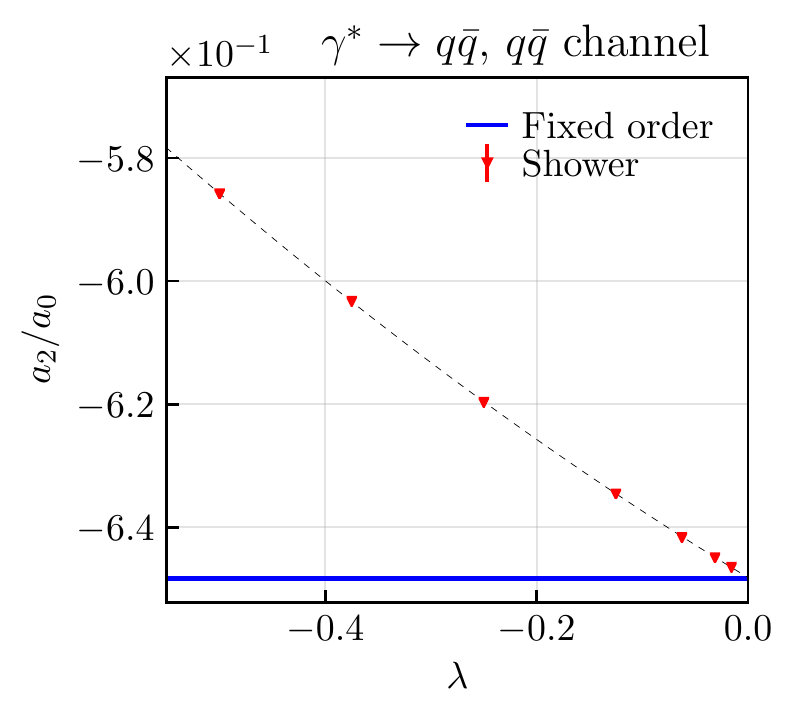}
  \end{subfigure}
    \caption{The ratio $a_2/a_0$ as a function of $\lambda =\alpha_s
      L$ for $\alpha_s = 10^{-7}$. In red we show the PanGlobal shower
      with $\beta=0$ and in blue the (integrated) analytic fixed-order result,
      see \cref{eq:fixed-order-slice-1}. The three panels show all channels
      (upper), the channel where a $g\rightarrow gg$ splitting is
      found inside the slice (lower left) and the channel where a
      $g\rightarrow q\bar{q}$ splitting is found inside the slice
      (lower right). The size of the modulation $a_2/a_0$ decreases
      with $\lambda$ in all three plots, and approaches the fixed
      order result for $\lambda\rightarrow 0$.  We use the
      leading-$\nc$ approximation with
      $C_F=\frac{C_A}{2}=\frac{3}{2}$.}
    \label{fig:a2a0vslambda}
\end{figure}

To provide reference results for future studies, in
\cref{tab:lambda-scan} we consider three values of $\lambda = -0.5, -0.25$ and
$-0.125$ and show  the values of the coefficients $a_0$ and $a_2$, as defined in
\cref{eq:all-order-general-form}, and extracted through a
Fourier cosine transform.
The results are given  separately for the sum over flavour channels
and for the $g \to gg$ and $g \to q\bar q$ splitting channels.
The magnitude of the rest channel can be deduced from the difference
and has an $a_2$ value that is consistent with zero.
\begin{table}[]
   \centering
{\footnotesize
   \begin{tabular}{llll}
     \toprule
    \multicolumn{1}{l|}{$\lambda$} & $-0.5$ & $-0.25$ & $-0.125$ \\ \midrule \toprule  \multicolumn{1}{c|}{}           & \multicolumn{3}{l}{All channels} \\ \midrule \multicolumn{1}{l|}{$a_0$}     & $+9.63 \cdot 10^{-2}$ & $+2.80 \cdot 10^{-2}$ & $+7.34 \cdot 10^{-3}$  \\ \multicolumn{1}{l|}{$a_2$}     & $-2.68 \cdot 10^{-3}$ & $-8.30 \cdot 10^{-4}$ & $-2.24 \cdot 10^{-4}$  \\ \multicolumn{1}{l|}{$a_2/a_0$}     & $-2.78 \cdot 10^{-2}$ & $-2.96 \cdot 10^{-2}$ & $-3.05 \cdot 10^{-2}$
\end{tabular}}
{\footnotesize
   \begin{tabular}{llll}
     \toprule
    \multicolumn{1}{c|}{}           & \multicolumn{3}{l}{$gg$} \\ \midrule \multicolumn{1}{l|}{$a_0$}     & $+8.33 \cdot 10^{-2}$ & $+2.45 \cdot 10^{-2}$ & $+6.44 \cdot 10^{-3}$  \\ \multicolumn{1}{l|}{$a_2$}     & $+4.01 \cdot 10^{-3}$ & $+1.25 \cdot 10^{-3}$ & $+3.36 \cdot 10^{-4}$  \\ \multicolumn{1}{l|}{$a_2/a_0$}     & $+4.81 \cdot 10^{-2}$ & $+5.10 \cdot 10^{-2}$ & $+5.22 \cdot 10^{-2}$
\end{tabular}}
{\footnotesize
   \begin{tabular}{llll}
     \toprule
    \multicolumn{1}{c|}{}           & \multicolumn{3}{l}{$q\bar q$} \\ \midrule \multicolumn{1}{l|}{$a_0$}     & $+1.14 \cdot 10^{-2}$ & $+3.35 \cdot 10^{-3}$ & $+8.82 \cdot 10^{-4}$  \\ \multicolumn{1}{l|}{$a_2$}     & $-6.69 \cdot 10^{-3}$ & $-2.08 \cdot 10^{-3}$ & $-5.60 \cdot 10^{-4}$  \\ \multicolumn{1}{l|}{$a_2/a_0$}     & $-5.86 \cdot 10^{-1}$ & $-6.20 \cdot 10^{-1}$ & $-6.35 \cdot 10^{-1}$
\end{tabular}}
\caption{Numerical values of the coefficients $a_0$, $a_2$, as defined
  in Eq.~(\ref{eq:all-order-general-form}), and the
  size of the spin correlations $a_2/a_0$ for the observable $\dpsisl$
  ($y_{\text{max}}=1,\,z_{\text{cut}}=0.1$), as extracted by a
  Fourier cosine transform from the PanGlobal $\beta=0$ shower with
  soft corrections included, for values of $\lambda \equiv \as L \in
  \lbrace -0.5,-0.25,-0.125 \rbrace$. The contributions
  are given separately for each flavour channel, and their statistical
  uncertainty is at most one in the last quoted digit for all results.
  We employ the leading-$\nc$ approximation with
  $C_F=\frac{C_A}{2}=\frac{3}{2}$.}
\label{tab:lambda-scan}
\end{table}
The ratio $a_2/a_0$ is furthermore shown in
\cref{fig:a2a0vslambda} as a function of $\lambda$, where the
fixed-order result is also included, see \cref{eq:fixed-order-slice-1}.  The size of the modulation
$a_2/a_0$ decreases approximately linearly with $\lambda$ in all three
figures, and approaches the fixed-order result for $\lambda\rightarrow
0$ as expected.
A fit of the points confirms that there are also non-linear terms
present, with small numerical coefficients.

\section{Conclusions}
\label{sec:concl}

We have demonstrated in this article that it is relatively
straightforward to extend the Collins-Knowles algorithm for collinear
spin correlations so as to address also the spin correlations of soft
emissions, in the leading-$\nc$ limit.
Within the PanScales shower framework, this was the last step needed
to obtain massless final-state showers that fully satisfy the
PanScales NLL conditions at leading-$\nc$.
In particular it is critical for reproducing the correct azimuthal
structure of matrix elements of nested sequences of soft and then
collinear splittings, regardless of the angle of the soft splitting.
Within the frame of an individual dipole, the structure of the spin
correlations for soft emissions is remarkably simple, as it has to be
given the invariance of soft emission with respect to boosts along the
parent dipole direction, cf.\ Fig.~\ref{fig:results-alphas2}
(right).\footnote{That simplicity could, conceivably, also be
  exploited directly in formulating parton shower algorithms with soft
  spin correlations, though we envisage that this would require more
  gymnastics in transporting reference azimuthal angles from one
  dipole frame to another.}
A purely collinear implementation of the spin correlations, e.g.\ that
of our earlier work~\cite{Karlberg:2021kwr}, can alter that simple
structure with $\order{1}$ relative artefacts at angles commensurate
with parent-dipole opening angles.

Beyond the leading-$\nc$ limit, it would no longer be sufficient to
consider a single parent dipole for any given large-angle soft
emission.
This would complicate the treatment of soft spin correlations in the
same way that it complicates the treatment of soft emission more
generally.
In principle one could adapt the NODS colour treatment of
Ref.~\cite{Hamilton:2020rcu} to also address spin correlations up to
some fixed order.
However, we leave this to future work, especially in view of the
observation in Appendices~\ref{app:analytic-mes} and \ref{sec:app:FC}
that the original spin-agnostic NODS approach, combined with our
leading-$\nc$ soft spin correlations, already reproduces the
full-colour $3$-emission matrix elements' azimuthal modulation to
within a few percent.

As was the case until recently~\cite{Chen:2020adz,Karlberg:2021kwr}
also for collinear spin correlations, there are, to our knowledge, no
standard observables geared to the measurement of soft spin
correlations.
The observable that we introduce in \cref{sec:new-observable}
addresses this gap.
Our showers provide
reference resummations for this observable, cf.\
Figs.~\ref{fig:allorder-dpsi} and \ref{fig:a2a0vslambda} and
Table~\ref{tab:lambda-scan}.

Overall, spin correlations in the soft limit lead to significant
azimuthal modulations, and may be important also in work towards
higher-order parton showers (see, for example, the discussion in
Ref.~\cite{Gellersen:2021eci}).
We hope that our results can pave the way to their straightforward
inclusion in a range of parton showers.

\section*{Acknowledgements}

We are grateful to our PanScales collaborators (Melissa van Beekveld,
Mrinal Dasgupta, Fr\'ed\'eric Dreyer, Basem El-Menoufi, Silvia Ferrario
Ravasio, Rok Medves, Pier Monni, Gr\'egory Soyez and
Alba Soto Ontoso), for their work on the code, the underlying
philosophy of the approach and comments on this manuscript.

This work was supported
by a Royal Society Research Professorship
(RP$\backslash$R1$\backslash$180112) (GPS, LS),
by the European Research Council (ERC) under the European Union’s
Horizon 2020 research and innovation programme (grant agreement No.\
788223, PanScales) (AK, GPS, KH, RV), 
by the Science and Technology Facilities Council (STFC) under
grants ST/T000856/1 (KH) and ST/T000864/1 (GPS),
by Linacre College (AK) and by Somerville College (LS).

\appendix

\input{analytic_me.tex}

\section{Fixed-order tests at full colour}
\label{sec:app:FC}
\begin{figure}
  \centering
  \includegraphics{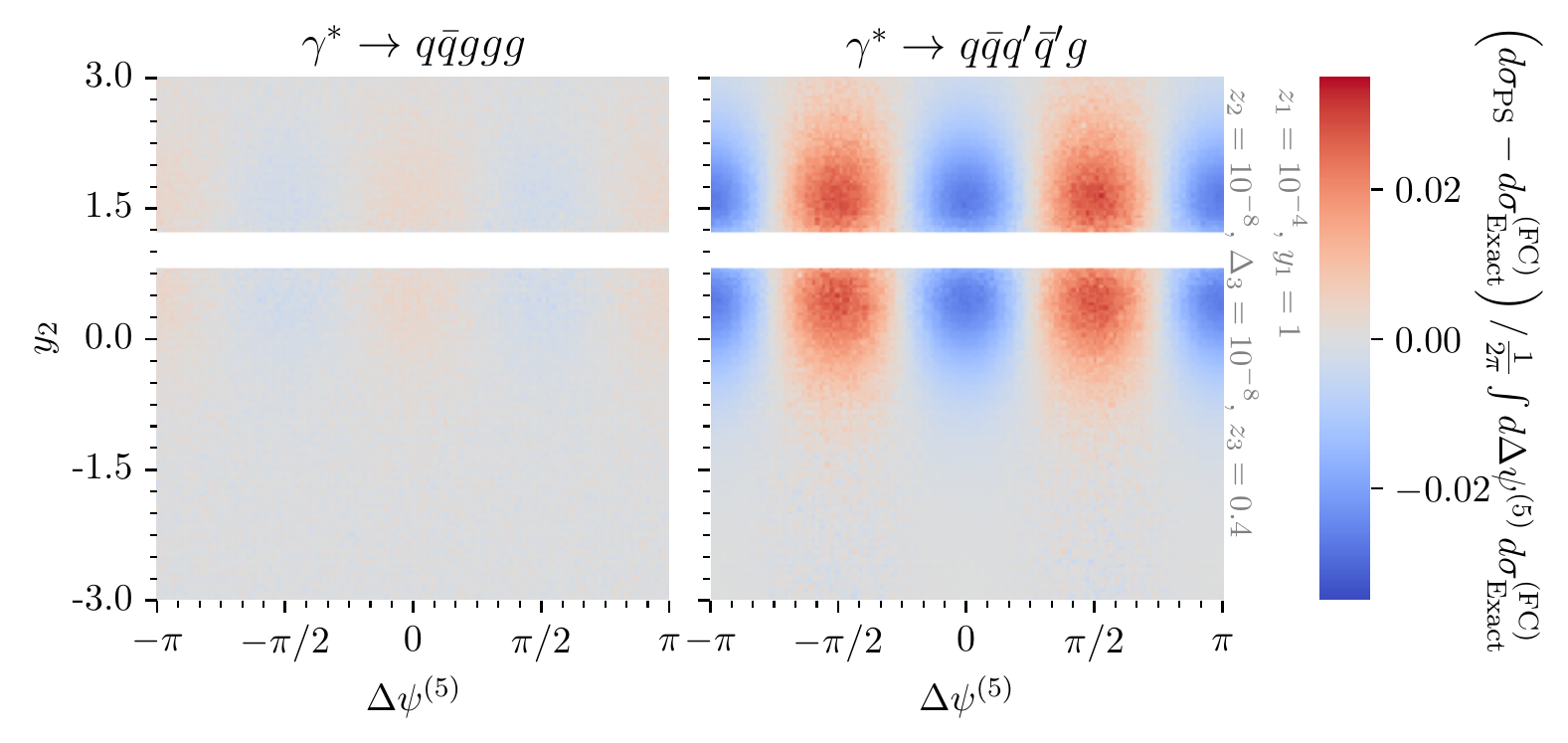}
  \caption{The ratio
    $\left(\text{d}\sigma_{\text{PS}}-\text{d}\sigma_{\text{Exact}}^\text{(FC)}\right)
    /\frac{1}{2\pi}\int{\Delta\psi^{(5)}\text{d}\sigma_{\text{Exact}}^\text{(FC)}}$ of
    the difference between the parton shower and the correct squared
    full colour tree-level matrix element, for $e^+e^- \to q\bar q
    ggg$ (left) and $e^+e^- \to q\bar q q' \bar q' g$ (right), for the
    azimuthal correlation between the primary and secondary splitting
    planes $\Delta \psi^{(5)}$. The parton shower result is shown
    using the NODS method of Ref.~\cite{Hamilton:2020rcu}.}
    \label{fig:fullcolour}
\end{figure}
\begin{figure}
  \centering
  \includegraphics[page=2]{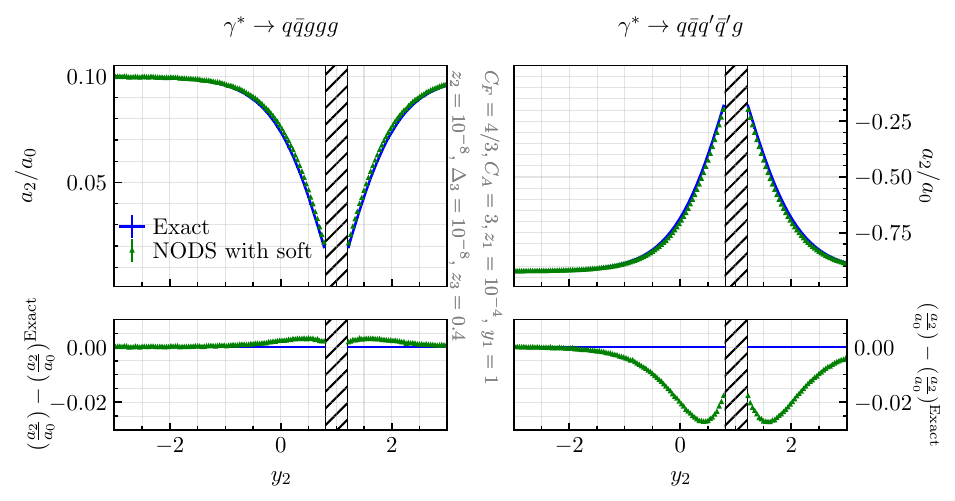}
  \caption{The ratio $a_2/a_0$ extracted from the double-differential
    cross section in \cref{fig:fullcolour}, as a function of $y_2$,
    comparing the NODS-enhanced soft-spin algorithm with the
    full-colour (squared) matrix-element results from
    Appendix~\ref{app:analytic-mes}.}
    \label{fig:fullcolour-1d}
\end{figure}

In the main body of the paper, we have considered only the leading-colour approximation by
setting $C_F = C_A/2 = 3/2$, as we do not expect our implementation of spin
correlations to reproduce the full-colour structure at NLL.

Here we consider the performance of our method beyond the
leading-$\nc$ limit, in the context of 
the NODS (nested ordered double-soft) scheme for including subleading-colour
effects in parton showers at leading-logarithmic level, as introduced in
Ref.~\cite{Hamilton:2020rcu}.\footnote{Alternative methods
for the inclusion of subleading-colour effects in parton showers
can be found in Refs.~\cite{Platzer:2012np,Nagy:2012bt,Nagy:2015hwa,Platzer:2018pmd,Nagy:2019pjp,Forshaw:2019ver,DeAngelis:2020rvq,Hoche:2020pxj,Holguin:2020oui}.}
The NODS scheme consists of a local (squared) matrix-element
correction, which ensures that the shower reproduces the correct
full-colour radiation pattern for every pair of soft energy-ordered
commensurate-angle gluons, as long as other emissions are
well-separated from that pair, in rapidity.
In the tests that we show here, we use the large-$\nc$ soft-spin
approach of \cref{sec:technical}, multiplied by the
spin-averaged subleading-$\nc$ NODS correction factors.

In \cref{fig:fullcolour,fig:fullcolour-1d}, we compare the full-colour matrix
element against the dipole version of the PanLocal shower (with $\beta=0.5$)
using the NODS procedure, for the same 5-parton configurations as those
presented in \cref{sec:fo-validation}. Note that the results presented here
are independent of the exact PanScales shower choice.
For a soft gluon emission at rapidities close to the (fixed) first
gluon rapidity, $y_2 \sim y_1$, the parton-shower result does
not reproduce the correct azimuthal dependence at full colour, though
it does generate the correct azimuth-integrated normalisation. The
residual departures from the correct modulations
$(a_2/a_0)^\text{Exact}$, of the order of a few permille in the $gg$
channel and $3\%$ in the $q\bar q$ channel, are compatible with a
$1/\nc^2$ correction to the leading-colour modulations seen in
Fig.~\ref{fig:results-alphas3}.

\logbook{76dacefb}{see plots for $a_0$, $a_2$ in
2020-eeshower/analyses/soft-spin-alphas3-ME/mevsps-5-body-eta_FC-Keith_1d.pdf.
The quantities plotted in 2D above correspond to
$\frac{a_0^\text{PS}-a_0^\text{ME}}{a_0^\text{ME}} +
\frac{a_2^\text{PS}-a_2^\text{ME}}{a_0^\text{ME}} \text{cos} 2\Delta \psi$ (i.e.
the $\lesssim 1\%$ in $gg$, $\sim 3\%$ in $q \bar q$ can be read directly off
for NODS). No higher-order cosine terms (the fact that segment looks non-cosine
like on the 2D plots in the paper is because the normalisations $a_0$ don't
agree).}

While these residual effects are numerically small, it could still be of
conceptual interest to attempt to extend the NODS procedure to work at
amplitude level, so as to obtain the correct full-$\nc$ spin
correlations for commensurate-angle energy-ordered soft pairs.
We leave the development and study of such an approach to potential
future work.

\section{Sensitivity to choice of reference vector} \label{sec:app:reference-direction}

\begin{figure}
\centering
\includegraphics[page=2]{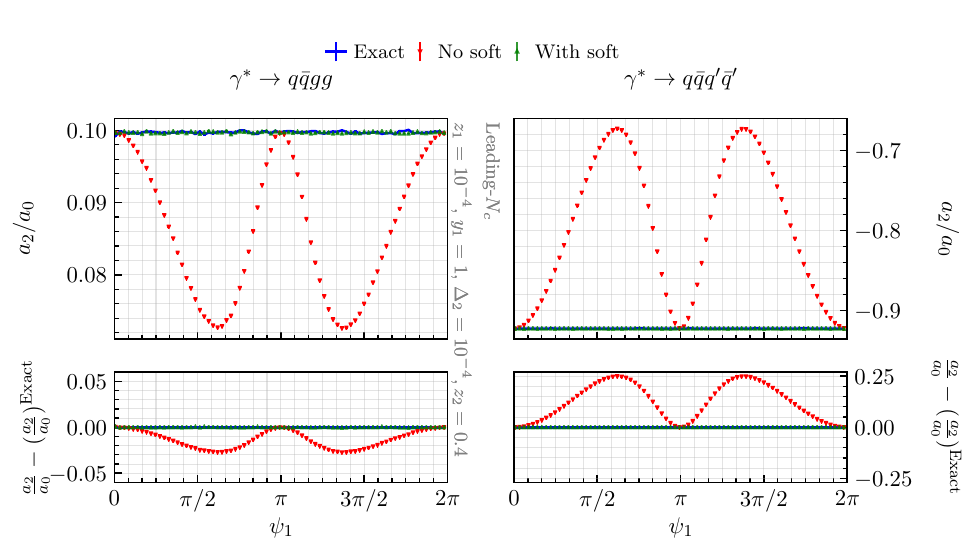}
\caption{The same matrix element comparison as \cref{fig:results-alphas2-FT},
but as a function of the intermediate gluon's azimuthal angle $\psi_1$ instead
of its rapidity $y_1$. The latter is fixed to $y_1=1$.
}
\label{fig:reference-direction-dependence}
\end{figure}

The branching amplitudes defined in \cref{eq:branch-amp-spinor-prod,eq:branch-amp-spinor-prod-soft}
involve spinor products, which must be evaluated numerically.
Without loss of generality, the spinor product may be expressed as 
\begin{equation}
S_{+}(p_b, p_c) = \frac{1}{\sqrt{p_b{\cdot}k_0} \sqrt{p_c{\cdot}k_0}} \bigg[ (p_c{\cdot}k_0) (p_b{\cdot}k_1) - (p_b{\cdot}k_0)(p_c{\cdot}k_1) - i \epsilon_{\mu \nu \alpha \beta} k_0^{\mu} k_1^{\nu} p_b^{\alpha} p_c^{\beta} \bigg],
\end{equation} 
where $k_0$ and $k_1$ are arbitrary reference vectors which obey $k_0^2 = k_0{\cdot}k_1 = 0$ and $k_1^2 = -1$.
In the evaluation of a complete, gauge-invariant squared scattering amplitude, any dependence on these reference directions must necessarily vanish.
However, the branching amplitudes used in the Collins-Knowles algorithm only reproduce the full scattering amplitude in the relevant singular limits.
Outside of this limit, a spurious dependence remains.
In particular, in Ref.~\cite{Karlberg:2021kwr} it was shown that this dependence indeed vanishes in the collinear limit,
but in that implementation it remains in the soft limit. 
Furthermore, because a definite choice must be made, this effect depends on the event orientation.

This issue is illustrated in \cref{fig:reference-direction-dependence}, where the second-order matrix element comparison of \cref{fig:results-alphas2-FT}
is repeated, but now as a function of $\psi_1$, the azimuth of the soft gluon emission, instead of $y_1$ which is fixed to $y_1=1$. 
As explained, without soft corrections the result depends on $\psi_1$, only reproducing the soft matrix element for the specific values $\psi_1 = \{0, \pi, 2\pi\}$.
When soft corrections are instead enabled, the Collins-Knowles algorithm reproduces the soft matrix element for all values of $\psi_1$.

\bibliographystyle{JHEP}
\bibliography{MC}

\end{document}

%% file: analytic_me.tex
\section{Analytic matrix elements}
\label{app:analytic-mes}

We write out below the matrix elements calculated analytically for emissions
from the Born $\gamma^* \to q\bar q$ at second and third order, which are used
in the comparisons performed in \cref{sec:fo-validation}.
We will use the following auxiliary functions for the different emission terms:
\begin{align}
\mathcal{A}_3(a,b,c)= & \frac{4s_{a,c}}{s_{a,b}s_{b,c}}\,,\\
\mathcal{A}_4(a,b,c,d)= &
\frac{8s_{a,d}^{2}}{s_{a,b}s_{c,d}\left(s_{a,b}+s_{a,c}\right)\left(s_{b,d}+s_{c,d}\right)}+\frac{8s_{a,d}}{s_{a,b}s_{b,c}}\left(\frac{1}{s_{c,d}}+\frac{1}{s_{b,d}+s_{c,d}}\right)
\nonumber \\ + &
\frac{8s_{a,d}}{s_{b,c}\left(s_{a,b}+s_{a,c}\right)}\left(\frac{1}{s_{c,d}}-\frac{4}{s_{b,d}+s_{c,d}}\right)
\nonumber \\+ &
\frac{2}{s_{b,c}^{2}}\left(\frac{s_{a,b}-s_{a,c}}{s_{a,b}+s_{a,c}}-\frac{s_{b,d}-s_{c,d}}{s_{b,d}+s_{c,d}}\right)^{2}\,,\\
\delta(q,\bar q\,;\,a,b,c)\,=\, & \mathcal{A}_4(q,c,b,\bar
q)\,+\,\mathcal{A}_3(q,b,\bar q)\,\mathcal{A}_3(q,c,a) \nonumber\\ +\, &
\mathcal{A}_3(q,c,\bar
q)\,\mathcal{A}_3(q,b,a)-\,\frac{1}{2}\,\mathcal{A}_3(q,b,\bar
q)\,\mathcal{A}_3(q,c,\bar q)\,,
\end{align}
with the Lorentz invariant $s_{i,j} = 2 p_i \cdot p_j$. Furthermore, we will
express the matrix elements for final states containing a quark pair $q',\,\bar
q'$ below, with the help of the following functions:
\begin{align}
\mathcal{B}_2(a,b\,;\,q',\bar q') = & \frac{2}{s_{q',\bar
q'}}\,\bigg(\frac{s_{a,b}}{\left(s_{a,q'}+s_{a,\bar
q'}\right)\left(s_{b,q'}+s_{b,\bar q'}\right)} \nonumber \\
-&\frac{1}{4s_{q',\bar q'}}\left(\frac{s_{a,q'}-s_{a,\bar
q'}}{s_{a,q'}+s_{a,\bar q'}}-\frac{s_{b,q'}-s_{b,\bar q'}}{s_{b,q'}+s_{b,\bar
q'}}\right)^{2}\,\bigg)\,, \\ \mathcal{B}_3(a,b,c\,;\,q',\bar q')\,=\, &
\frac{1}{s_{a,b}}\,\bigg(\,\big(\frac{s_{a,b}}{(s_{a,q'}+s_{a,\bar
q'})(s_{b,q'}+s_{b,\bar q'})}+\frac{s_{a,c}}{(s_{a,q'}+s_{a,\bar
q'})(s_{c,q'}+s_{c,\bar q'})} \nonumber \\-&\frac{s_{b,c}}{(s_{b,q'}+s_{b,\bar
q'})(s_{c,q'}+s_{c,\bar q'})}\big) \nonumber \\-&\frac{1}{2s_{q',\bar
q'}}\big(\frac{s_{a,q'}-s_{a,\bar q'}}{s_{a,q'}+s_{a,\bar
q'}}-\frac{s_{b,q'}-s_{b,\bar q'}}{s_{b,q'}+s_{b,\bar
q'}}\big)\,\big(\frac{s_{a,q'}-s_{a,\bar q'}}{s_{a,q'}+s_{a,\bar
q'}}-\frac{s_{c,q'}-s_{c,\bar q'}}{s_{c,q'}+s_{c,\bar q'}}\big)\,\bigg)\,.
\end{align}

\subsection*{The 4-parton matrix element for $\gamma^* \to q\bar q g_1 g_2$}

We assume the following ordering in the energies of the final-state gluons:
\begin{itemize}
\item the energy of the two gluons $g_1$ and $g_2$ is much smaller than the
energies of the Born (anti-)quark, $E_{g_1}, E_{g_2} \ll E_q, E_{\bar q}$ (no ordering
is assumed on the relative energy of the two gluons).
\end{itemize}
The matrix element at leading colour is given by
\begin{align}
\lim_{\substack{\nc\rightarrow\infty }
}\,\frac{\left|\mathscr{M}_{q\bar{q}g_{1}g_{2}}\right|^{2}}{\left|\mathscr{M}_{q\bar{q}}\right|^{2}}
& =\ \left(4\pi\alpha_{{\scriptscriptstyle
S}}\right)^{2}\left(\frac{\nc}{2}\right)^{2}\big(\mathcal{A}_4(q,g_1,g_2,\bar
q) + \mathcal{A}_4(q,g_2,g_1,\bar q) \big)\,.
\end{align}

\subsection*{The 5-parton matrix element for $\gamma^* \to q\bar q g_1 g_2 g_3$}

We assume the following ordering in the energies of the final-state gluons:
\begin{itemize}
\item the energy of the first gluon $g_1$ is much smaller than the energies of
the Born quarks, $E_{g_1} \ll E_q, E_{\bar q}$
\item the energies of the second and third gluons, $g_2$ and $g_3$, are much
smaller than the energy of the first gluon, $E_{g_2}, E_{g_3} \ll E_{g_1}$ (no
ordering is assumed on the relative energy of $g_2$ and $g_3$).
\end{itemize}
We first give the matrix element at leading colour,
\begin{align}
\lim_{\substack{\nc\rightarrow\infty }
}\,\frac{\left|\mathscr{M}_{q\bar{q}g_{1}g_{2}g_{3}}\right|^{2}}{\left|\mathscr{M}_{q\bar{q}}\right|^{2}}
= &\ \left(4\pi\alpha_{{\scriptscriptstyle
S}}\right)^{3}\left(\frac{\nc}{2}\right)^{3} \mathcal{A}_3(q,g_1,\bar q) \nonumber \\
\times  &\  \left(
              \mathcal{A}_3(q,g_3,g_1) \mathcal{A}_3(g_1,g_2,\bar q) +
              \mathcal{A}_4(q,g_2,g_3,g_1) +
              \mathcal{A}_4(g_1,g_2,g_3,\bar q)
            \right) \nonumber \\
     +  &\  \{ g_{2} \leftrightarrow g_{3} \}\,.
\end{align}
At full colour, $N_C=3$, the matrix element is corrected to:
\begin{align}
\frac{\left|\mathscr{M}_{q\bar{q}g_{1}g_{2}g_{3}}\right|^{2}}{\left|\mathscr{M}_{q\bar{q}}\right|^{2}}\,=\,
& \left(\frac{2C_{F}}{C_{A}}\right)\,\lim_{\substack{N_{{\scriptscriptstyle
C}}\rightarrow\infty }
}\,\frac{\left|\mathscr{M}_{q\bar{q}g_{1}g_{2}g_{3}}\right|^{2}}{\left|\mathscr{M}_{q\bar{q}}\right|^{2}}\,\nonumber
\\
- & \left(\frac{2C_{F}}{C_{A}}\right)\,\left(4\pi\alpha_{{\scriptscriptstyle
  S}}\right)^{3}\,\frac{\nc}{8}\,\mathcal{A}_3 (q,g_{1},\bar
q)\,\left(\delta(q,\bar q\,;\,g_1,g_2,g_3)\,+\delta(\bar
q,q\,;\,g_1,g_2,g_3)\,\right)\,\nonumber \\ + &
\left(\frac{2C_{F}}{C_{A}}\right)\,\left(4\pi\alpha_{{\scriptscriptstyle
S}}\right)^{3}\,\frac{1}{8\nc}\,\mathcal{A}_3(q,g_{1},\bar
q)\,\mathcal{A}_3(q,g_{2},\bar q)\,\mathcal{A}_3(q,g_{3},\bar q)\,.
\label{eq:me-qqggg-fullcolour}
\end{align}
The matrix element given in Eq.~(\ref{eq:me-qqggg-fullcolour}) is the one used
in the comparisons to the NODS scheme presented in
Appendix~\ref{sec:app:FC}.

\subsection*{The 4-parton matrix element for $\gamma^* \to q\bar q q' \bar q'$}

We assume the following ordering in the energies of the final-state quarks:
\begin{itemize}
\item the energy of the two quarks $q'$ and $\bar q'$ is much smaller than the
energies of the Born quarks, $E_{q'}, E_{\bar q'} \ll E_q, E_{\bar q}$.
\end{itemize}
The matrix element, exact in $N_C$, is given by
\begin{align}
\frac{\left|\mathscr{M}_{q\bar{q}q'\bar
q'}\right|^{2}}{\left|\mathscr{M}_{q\bar{q}}\right|^{2}}\,= &
\left(4\pi\alpha_{s}\right)^{2}\,4n_{f}T_{R}C_{F}\,\mathcal{B}_2(q,\bar
q\,;\,q',\bar q')\,.
\end{align}

\subsection*{The 5-parton matrix element for $\gamma^* \to q\bar q q' \bar q' g$}

We assume the following ordering in the energies of the final-state particles:
\begin{itemize}
\item the energy of the gluon $g$ is much smaller than the energies of the Born
quarks, $E_{g} \ll E_q, E_{\bar q}$.
\item the energy of the two quarks $q'$ and $\bar q'$ is much smaller than the
energy of the gluon, $E_{q'}, E_{\bar q'} \ll E_{g}$
\end{itemize}
The matrix element, exact in $N_C$, is given by
\begin{align}
\frac{\left|\mathscr{M}_{q\bar{q}q'\bar
q'g}\right|^{2}}{\left|\mathscr{M}_{q\bar{q}}\right|^{2}}
=&
\left(4\pi\alpha_{s}\right)^{3}\,4n_{f}T_{R}\,\left(\frac{N_C}{2}\right)^{2}\,\mathcal{A}_3(q,g,\bar q) \nonumber \\ \times &
\bigg(\mathcal{B}_2(q,g\,;\,q',\bar q')-\frac{1}{N_C^{2}}\left(4\mathcal{B}_3(q,g,\bar
q\,;\,q',\bar q')+\mathcal{B}_3(g,q,\bar q\,;\,q',\bar q')-\mathcal{B}_2(q,\bar q\,;\,q',\bar q')\right)\nonumber \\ +&
\frac{1}{2N_C^{4}}\left(4\mathcal{B}_3(q,g,\bar q\,;\,q',\bar q')-\mathcal{B}_2(q,\bar
q\,;\,q',\bar q')\right)\bigg)\nonumber \\ + & \left\{ q\leftrightarrow\bar q\right\} \,.
\label{eq:me-qqqqg-fullcolour}
\end{align}
The full-colour matrix element in Eq.~(\ref{eq:me-qqqqg-fullcolour}) is used in
the comparisons to the NODS results in Appendix~\ref{sec:app:FC}. In
the limit $N_C \to \infty$, only the first term in the parenthesis remains.
